\documentclass[aps,pra,reprint,superscriptaddress]{revtex4-1}

\usepackage{amsmath}
\usepackage{amssymb}
\usepackage{amsfonts}

\usepackage{graphicx}
\newcommand{\da}{^\dagger}

\newcommand{\ddx}[1]{{\partial_{#1}}}

\newcommand{\dfdx}[2]{\dfrac{\partial#1}{\partial#2}}

\newcommand{\br}[1]{\langle #1 \vert}
\newcommand{\ke}[1]{\vert #1 \rangle}

\newcommand{\ev}[1]{\langle #1 \rangle}

\usepackage{xcolor}

\begin{document}
\title{Quantum Control with Quantum Light of Molecular Nonadiabaticity}

\author{Andr\'as Csehi}
\affiliation{Department of Theoretical Physics, University of Debrecen, H-4002 Debrecen, PO Box 400, Hungary}
\affiliation{ELI-ALPS, ELI-HU Non-Profit Ltd, H-6720 Szeged, Dugonics t\'er 13, Hungary}
\author{G\'abor J. Hal\'asz}
\affiliation{Department of Information Technology, University of Debrecen, H-4002 Debrecen, PO Box 400, Hungary}
\author{\'Agnes Vib\'ok}
\affiliation{Department of Theoretical Physics, University of Debrecen, H-4002 Debrecen, PO Box 400, Hungary}
\affiliation{ELI-ALPS, ELI-HU Non-Profit Ltd, H-6720 Szeged, Dugonics t\'er 13, Hungary}
\author{Markus Kowalewski}
\affiliation{Department of Physics, Stockholm University, AlbaNova University
Centre 106 91 Stockholm, Sweden}
\email{markus.kowalewski@fysik.su.se}

\date{\today}%

\begin{abstract}
Coherent control experiments in molecules are often done with shaped laser fields.
The electric field is described classically and control
over the time evolution of the system is achieved by shaping
the laser pulses in the time or frequency domain. Moving on
from a classical to a quantum description of the light
field allows to engineer the quantum state of light to steer chemical processes.
The quantum field description of the photon mode allows
to manipulate the light-matter interaction directly in phase-space.
In this paper we will demonstrate the basic principle of coherent
control with quantum light on the avoided crossing in lithium fluoride.
Using a quantum description of light together with the nonadiabatic
couplings and vibronic degrees of freedoms opens up new perspective on quantum control.
We show the deviations from control with purely classical light field
and how back-action of the light field becomes important in a few photon regime.
\end{abstract}

\maketitle

\section{Introduction}

Coherent control \cite{Warren93,Bartanaa01cp,vdHoff12pccp,Koch12cr,Dantus04cr,Shapiro,Brif10njp} has greatly contributed to the understanding
of how photo-chemical reactions can be manipulated and what the limits of controllability are.
In a typical optimal control experiment a short laser pulse drives optical or infrared transitions
aiming at optimizing a specific objective such as the yield of a photo-chemical reaction.
This can be achieved by creating interference between light induced pathways \cite{Brumer86cpl,Shapiro88cpl}
or by steering wave packets in a desired direction \cite{Tannor86jcp,Gordon97arpc}. These control principles have been realized
in optimal control experiments and investigated theoretically by means of optimal
control theory.
Given an input laser pulse of a fixed temporal length one can then shape the pulse in the
frequency domain by changing phase, amplitude, and polarization of the frequency components in the pulse
spectrum. Thus in a classical description of light there are three variables for a single frequency mode.
However, in a quantum description of light the behavior of a single frequency mode can be described by
a variable number of Fock-states, their amplitudes and phase (and polarization). This new description leads
to a wealth of new control knobs for coherent control.
The quantum nature of light becomes relevant in the few-photon regime.
This regime can be reached either with low intensity beams or in a spatially confined field mode,
such as in a nano-cavity.

In the latter situation the strong light-matter coupling
can be achieved by considering the molecules to interact with a confined
light mode of the microscale or nanoscale optical cavities \cite{Aspelmeyer}.
Such hybrid light-matter systems are then characterized by the properties of
the common light and matter eigen state and are called polaritons or dressed states.

Over the past few years, polaritonic chemistry became an emerging
field which provides a novel tool for modifying and controlling the
chemical structure and dynamics. Several experimental \cite{Ebbesen1,Ebbesen2,Ebbesen3,Ebbesen4,Ebbesen5,Ebbesen6}
and theoretical \cite{Feist1,Feist2,Feist3,Feist4,Feist5,Herrera1,Herrera2,Herrera3,Joel1,Joel2,Joel3,Markus1,Kowalewski16jcp,Markus3,Vibok1,Vibok2,Rubio1,Rubio2,Rubio3,Flick1,Triana18jpca,Oriol1,triana19prl}
activities are concentrated in this field since the pioneering experimental
work by the group of Ebbesen, when it was observed that the strong
light-matter coupling could change the chemical landscapes and chemical
reaction \cite{Ebbesen1}. Among others it was found that the strong
coupling can modify the absorption spectra \cite{Ebbesen2,Ebbesen4,Feist1,Vibok2},
the nonadiabatic dynamics \cite{Markus1,Kowalewski16jcp,Markus3}, the supermolecular
polaritonic states provide very fast non-radiative energy transfer
\cite{Ebbesen4}.

Coherent control with quantized light fields has been discussed from a fundamental
point of view in Refs. \cite{Shapiro,Sun16arxiv} and a generalized optimal control
approach based on a quantum description of light has been proposed by Gruebele \cite{Gruebel01cp}.
Explicit quantum light coherent control applications that have been
proposed include the control of qbits in ions chains \cite{Shapiro11prl},
control of two-photon transitions \cite{Schlawin17njp} in atoms,
and its application to spectroscopy \cite{Rahav10pra,Dorfman16rmp}.
In this paper we will discuss the basic opportunities for coherent quantum control
that can be achieved with typical quantum states of light, such as Fock-states, squeezed states, and coherent states and apply it to control of a nonadiabatic coupling.
A study showing the general differences between quantum and classical light has been
presented in Ref.\ \cite{Triana18jpca}.
Here, we demonstrate how a single photon mode -- in quantum or classical description --
may be used to control the reaction outcome at the avoided crossing in LiF and present a general
coherent control concept for quantum light.
We will begin by presenting the underlying theoretical description of the coupled system of molecule and cavity, followed by an introduction of the envisioned control principle. Thereafter we will
present the results for the control of the nonadiabatic dynamics of the LiF molecule
and a discussion of the different scenarios.

\section{Theory}
\subsection{The Hamiltonian}
For the interaction of the quantized light field with a two-level system, we consider the full Rabi Hamiltonian \cite{Markus2,Schleich},
which is given by
\begin{align}\label{eq:H}
\hat{H}_{ec}&= H_e + H_c + H_I\\ &=\frac{\hbar\omega_0}{2}(2\hat{\sigma}^\dagger\hat{\sigma}-1)+\hbar\omega_c\hat{a}^\dagger\hat{a}
+\hbar g
(\hat{a}^\dagger+\hat{a})(\hat{\sigma}^\dagger+\hat{\sigma})\nonumber
\end{align}
where $H_e$, $H_c$ and $H_I$ describe the electronic and photon degrees of freedom, as well as
the light-matter interaction.
Here, $\sigma=\ke{g}\br{e}$ acts on the $\ke{g}$ electronic ground state and the $\ke{e}$ excited state,
$\hat{a}^{(\dagger)}$ is the bosonic annihilation (creation) operators of the photon mode,
$\hbar\omega_0=\hbar(\omega_e-\omega_g)$
is the energy difference between the electronic states, and $\omega_c$ is the resonance frequency of the
photon mode.
The vacuum Rabi frequency describing the light-matter coupling is:
\begin{align}
g=\dfrac{\mu_{eg} \varepsilon_c}{2\hbar}
\end{align}
and depends on the transition dipole moment $\mu_{ge}$ and on the vacuum field given by
\begin{equation}\label{eq:Ec}
\varepsilon_c = \sqrt{\frac{\hbar\omega_c}{V\epsilon_0}}\,,
\end{equation}
where $V$ is the quantization volume of the light mode.
In Eq.\ \ref{eq:H} we have kept the counter rotating terms $\sigma^\dagger\hat{a}^{\dagger}$ and
$\sigma\hat{a}$.
This is required to describe the ultra-strong coupling regime where $g$ is on the order of the
transition frequency $\omega_0$.

To allow for a convenient numerical description of the photon mode,
we use displacement coordinates rather than
the basis of Fock states. This can be achieved by expressing the
annihilation operator in terms of their photon displacement coordinates \cite{Schleich,Markus1}:
\begin{align}
a=\sqrt{\frac{\omega_c}{2\hbar}} \left( \hat{x} + \frac{i}{\omega_c} \hat{p} \right)\,,
\end{align}
with $\hat{p} = -i\hbar\ddx{x}$. The coordinate $x$ is a dimensionless coordinate
that is formally equivalent to a vibrational coordinate. The coupled Hamiltonian
from Eq.\ \ref{eq:H} then reads:
\begin{align}\label{eq:Hdirect}
H_{ec} &= \frac{\hbar\omega_0}{2} \left(2\hat{\sigma}\da \hat{\sigma} - 1\right) - \frac{\hbar^2}{2} \dfdx{^2}{x^2}
+\frac{1}{2}\omega_c^2 \hat{x}^2\\
&+ g\sqrt{2\hbar\omega_c} \hat{x} \left(\hat{\sigma}\da+\hat{\sigma} \right)\nonumber
\end{align}
For molecules, the transition frequency $\omega_0$ and the transition dipole moment $\mu_{ge}$  become quantities that
depend on the internuclear separation $R$ introducing nonadiabatic couplings \cite{Kowalewski16jcp}.
The total wave function is expanded in the adiabatic states
\begin{align}
    \Psi = \sum_k \psi_k(r;R) \phi_k(R,x)
\end{align}
where $r$ represents the electronic coordinates, $R$ is the internuclear distance and $k$ runs over the molecular electronic states (the $\Sigma_1$ ground and $\Sigma_2$ excited states of the LiF molecule are considered in the present work).
In the next step we combine Eq.\ \ref{eq:Hdirect} with the nuclear Hamiltonian in the basis
of the adiabatic states, which then reads:
\begin{align}\label{eq:Hkl}
H_{kl} &=  \delta_{kl} \left( -\dfrac{\hbar^2}{2m}\dfdx{^2}{R^2} + \hat{V}_{k}(R)
- \frac{\hbar^2}{2} \dfdx{^2}{x^2} +\frac{1}{2}\omega_c^2 \hat{x}^2 \right) \nonumber\\
&+ \left( 1 - \delta_{kl} \right) g(R) \sqrt{2\hbar\omega_c} \hat{x}\\
&+ \left( 1 - \delta_{kl} \right)\dfrac{1}{2m} \left( 2 f_{kl}(R) \dfdx{}{R} + \dfdx{}{R} f_{kl}(R) \right)\nonumber
\end{align}
where, $m$ is the reduced mass of the nuclei, $V_{k}(R)$ is the adiabatic potential energy curve of
the $k$-th electronic state,
and $\delta_{kl}$ is the Kronecker delta. The first-order nonadiabatic coupling matrix element
$f_{kl}(R)=\br{k} \partial_R \ke{l}$ describes the coupling at the avoided crossing (k,l=$\Sigma_1$, $\Sigma_2$).
For the sake of clarity and to demonstrate the
basic control possibility we neglect the diagonal dipole moments, which would cause couplings
between purely vibrational states.
In Eq.\ \ref{eq:Hkl}, the g(R) coupling strength is often expressed in terms of a parameter $\chi$ which is defined by the
relation $g(R)=\chi\cdot \mu_{kl}(R)\cdot \sqrt{\omega_c}$. This $\chi$ will be applied to characterize the coupling strength between the molecule and the photon mode.

By quantizating the light field, the state of the field is described by a
wave function rather than the wave form of the electric field.
The vibrational coordinate and the photon mode can now be treated on an equal footing.
The mode of the light field is treated like another vibrational mode with a harmonic potential.
In comparison the coupling term for the classical light-matter coupling is
\begin{align}\label{eq:HIclass}
    H_{I,class} = -\mu_{ge}E(t)
\end{align}
where $E(t)$ is the time-dependent electric field. The field properties of the
quantized photon mode and its time-dependence instead enter through the wave function rather than a Hamiltonian term such as Eq.\ \ref{eq:HIclass}.

\subsection{Nuclear Quantum Dynamics Simulations}

The MCTDH (multi configurational time-dependent Hartree) method \cite{mctdh1,mctdh2}
has been applied to solve the time-dependent Schr\"odinger-equation characterized by the Hamiltonian in Eq.\ \ref{eq:Hkl}.
The $R$ degree of freedom (DOF) was defined on a sin-DVR (discrete variable representation) grid
($N_R$ basis elements for $R = 0.846 - 21.16$\,\AA).
The photon mode, $x$ was described by $N_{x}$ Hermite-polynomials, $H_{m}(x)$
with $m=0,1,...,N_{x}-1$.
In the MCTDH wave function representation, these primitive basis sets ($\xi$) are then used to construct
the single particle functions ($\phi$) whose time-dependent linear combinations form the total nuclear wave packet ($\psi$)
\begin{equation}\label{eq3}
\begin{split}
   \phi^{(q)}_{j_{q}}(q,t)=\sum\limits_{i=1}^{N_q} c^{(q)}_{j_{q}i}(t)\xi^{(q)}_{i}(q)\ \ \ q=R,\ x \\
    \psi(R,x,t)=\sum\limits_{j_{R}=1}^{n_R} \sum\limits_{j_{x}=1}^{n_{x}}A_{j_{R},j_{x}}(t)\phi^{(R)}_{j_{R}}(R,t)\phi^{(x)}_{j_{x}}(x,t)
\end{split}
\end{equation}
The actual number of basis functions were $N_R=1069$ and $N_{x}=250 - 1550$ for the vibrational DOF and photon mode, respectively.
The number of single particle functions for both DOF and on both the $\Sigma_1$ and $\Sigma_2$ electronic states
were ranging from 10 to 44.
The values of $N_{x}$ and $n_{R} = n_{x}$ were chosen depending on the actual parameter values of the different quantum lights
so as to provide proper convergence.
In order to minimize unwanted reflections and transmissions caused by the finite length of the R-grid, complex absorbing potentials (CAP)
have been employed at the last $5.29$\,\AA\ of the grid.
The time of the propagation run was set $t_{final}$=200\,fs, hence the final $\Sigma_1$ state populations are calculated according to
\begin{equation}\label{eq4}
    P_{\Sigma_1}=\langle{\psi_{\Sigma_1}(R,x,t_{final})|\psi_{\Sigma_1}(R,x,t_{final}}\rangle
\end{equation}
The initial wave function $\psi(R,x,t=0)$ is a product of the electronic
wave function, the vibrational ground state, and one of the quantum light states described in
Eqs.\ \ref{eq:coh}, \ref{eq:sqvc}, or \ref{eq:sqcoh}:
\begin{align}
\psi(R,x,t=0) = \psi_{\Sigma_2} \otimes \psi_{v=0,\Sigma_1}(R) \otimes \Psi_{c/s/cs}(x)
\end{align}

\begin{figure}
\includegraphics[scale=0.5]{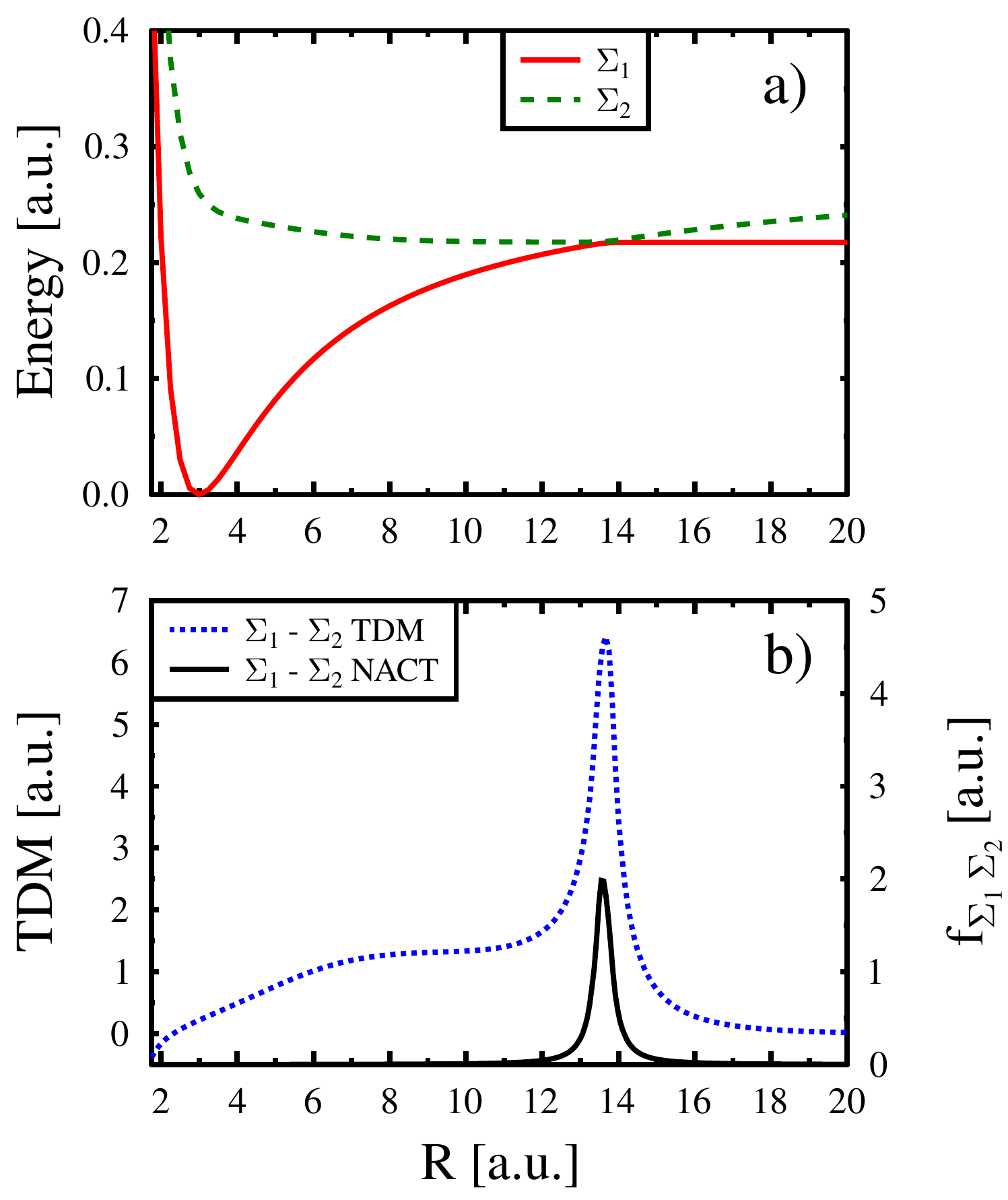}
\caption{(a) Potential energy curves of the $\Sigma_1$ (red solid line) and $\Sigma_2$ (green dashed line) electronic states of the LiF molecule applied in the present work. (b) The corresponding transition dipole moment and nonadiabatic coupling ($f_{\Sigma_1 \Sigma_2}$) curves are shown by blue dashed and black solid lines, respectively.}\label{fig:potentials}
\end{figure}

To calculate the potential energy, the dipole moment and the nonadiabatic coupling (NAC) curves of the LiF molecule,
the Molpro \cite{molpro} package has been utilized.
These quantities were calculated at the MRCI/CAS(6/12)/aug-cc-pVQZ level of theory.
In particular, $f_{\Sigma_1 \Sigma_2} (R)$ has been computed by finite differences of the MRCI electronic
wave functions. The number of active electrons and MOs in the
individual irreducible representations of the C$_{2v}$ point group were
A$_1$ $\rightarrow$ 2/5, B$_1$ $\rightarrow$ 2/3, B$_2$ $\rightarrow$ 2/3, A$_2$ $\rightarrow$ 0/1. The calculated electronic structure quantities shown in Fig.\ \ref{fig:potentials}

\subsection{Quantum States of Light}
In the following we introduce the quantum states of light that are used in the subsequent calculations.
Those states are used as initial states for light field at time $t=0$.
\subsubsection{Coherent state}
A coherent state is often regarded as the analog to a classical coherent light field.
The initial coherent state of the photon mode is given by a Gaussian \cite{quantopt},
\begin{align}
 \Psi_c(x) &= \left( \frac{\omega_c}{\pi\hbar} \right)^{1/4} \\
 &\times \exp \left[ -\left(\frac{x - \ev x_\alpha}{2\Delta x}\right)^2+\dfrac{\mathrm{i}}{\hbar}\ev p_\alpha\left(x-{\ev x_\alpha}\right)\right]\label{eq:coh}\nonumber\,,
\end{align}
where its parameters for width, initial displacement, and initial momentum are given by
\begin{align}
 \Delta x&=\sqrt{\dfrac{\hbar}{2\omega_c}}\\
 \ev x_\alpha &= \sqrt{\dfrac{\hbar\omega_c}{2}} \left ( \alpha + \alpha^* \right)\label{eq:evx}\\
 \ev p_\alpha &= -\mathrm{i} \sqrt{\dfrac{\hbar\omega_c}{2}} \left ( \alpha - \alpha^* \right)\label{eq:evp}\,.
\end{align}
The parameter $\alpha = |\alpha| \mathrm{e}^{\mathrm{i}\varphi}$ determines the amplitude of the displacement of the vacuum state. The phase $\varphi$ is its phase and corresponds to the carrier phase $\phi$ of a classical light field. The expectation value of the photon number is given by $\ev{n}=|\alpha|^2$.
An uncoupled coherent state oscillates back and forth along the photon displacement coordinate
(see Fig.\ \ref{fig:scheme}(a)) while keeping its width constant.

\subsubsection{Squeezed Vacuum State}
A squeezed vacuum state can be viewed as the ground state of a harmonic oscillator
with a modified width \cite{moller}:
 \begin{align}\label{eq:sqvc}
 \Psi_s(x) &= \left( \frac{\omega_c}{\pi\hbar} \right)^{1/4} \left(\cosh r + \mathrm{e}^{\mathrm{i}\theta} \sinh r \right)^{-1/2} \\
&\times \exp \left[ -\left(\frac{x}{2\Delta x}\right)^2\right] \nonumber\,,
 \end{align}
with the initial width
 \begin{align}
 \Delta x&=\sqrt{\dfrac{\hbar}{2\omega_c}}\left(\frac{\cosh r + \mathrm{e}^{\mathrm{i}\theta} \sinh r}{\cosh r - \mathrm{e}^{\mathrm{i}\theta} \sinh r}\right)^{1/2}\label{eq:Deltax}\,.
 \end{align}
Here $r$ is the squeezing parameter determining the extend of the squeezing and stretching
of the Gaussian. The phase $\theta$ is the squeezing phase and describes
whether the Gaussian is initial squeezed or stretched. Over time this state
will perform a "breathing motion" (see Fig.\ \ref{fig:scheme}(b)).
The average photon number of a squeezed state increases with the squeezing parameter: $\ev{n}=\sinh^2r$.

\subsubsection{Squeezed-Coherent State}
A squeezed-coherent state combines the idea of the squeezed vacuum state and a coherent state
and can be described by \cite{moller},
\begin{align}
 \Psi_{sc}(x) = &\left( \frac{\omega_c}{\pi\hbar} \right)^{1/4}
   \left(\cosh r + \mathrm{e}^{\mathrm{i}\theta} \sinh r \right)^{-1/2}\\
 &\times \exp \left[ -\left(\frac{x - \ev x_\alpha} {2\Delta x}\right)^2+\dfrac{\mathrm{i}}{\hbar}\ev p_\alpha\bigg(x-{\ev x_\alpha} \bigg)\right]\label{eq:sqcoh}\nonumber\,,
\end{align}
where $\Delta x$ is the same as in Eq.\ \ref{eq:Deltax}, and $\ev x_{\alpha}$ and $\ev p_{\alpha}$ are the same as in Eqs. \ref{eq:evx} and \ref{eq:evp}, respectively.
Its expectation value for the photon number is now determined by the displacement and the squeezing parameter: $\ev{n}=|\alpha|^2 + \sinh^2r $. Note that here both phases, $\theta$ and $\varphi$ determine the shape
of the initial wave packet.

\subsection{Quantum Control with Quantum Light}
The control scenario that we will compare in the following corresponds to a continuous
wave classical laser field. To demonstrate the basic principle and for the sake of clarity
we restrict the following discussion to a single mode.
In a single frequency laser field with a fixed frequency $\omega_L$
the two control parameters
available are amplitude $E_0$ and phase $\phi$ of the mode:
\begin{align}
E(t) = E_0 \cos(\omega_L t + \phi)
\end{align}
The quantum field mode introduced in Eq.\ \ref{eq:Hkl} replaces the classical field and
is now represented by a photon field wave function and its (uncoupled) eigen functions, the
eigen functions of the harmonic oscillator (or Fock-states).
The control variables are given by the initial state of the cavity mode
and thus constrained only by the size of its Hilbert space.
The interaction between two electronic states is then given by the operator
$g(R) \sqrt{2\hbar\omega_c} \hat{x}$ rather than $\mu_{ge} E(t)$ and is controlled
by the photon field wave function.
In contrast to a classical description of the electric field the molecule can now
also influence the state of the photon mode. This back-action will become
important in the few-photon regime and may create discrepancies between quantum and
classical description, which are otherwise expected to be equivalent.
Absorption and stimulated emission of single photons do not change the state of classical field. However, this assumption is only valid for large photon numbers.
In the limit of small photon numbers the exchange of photons between the
molecule and the field mode can significantly alter the state of the field mode.
The perfect Gaussian shape of a coherent state, for example, may end up severely distorted after interaction with the molecule (for an illustration of the dynamics in a simple atomic system see Figs.\ S4, S5, and S6 in the supplementary material).

The new control principles can now be explained in terms of the phase space of the
photon mode.
\begin{figure}
\includegraphics[scale=0.7]{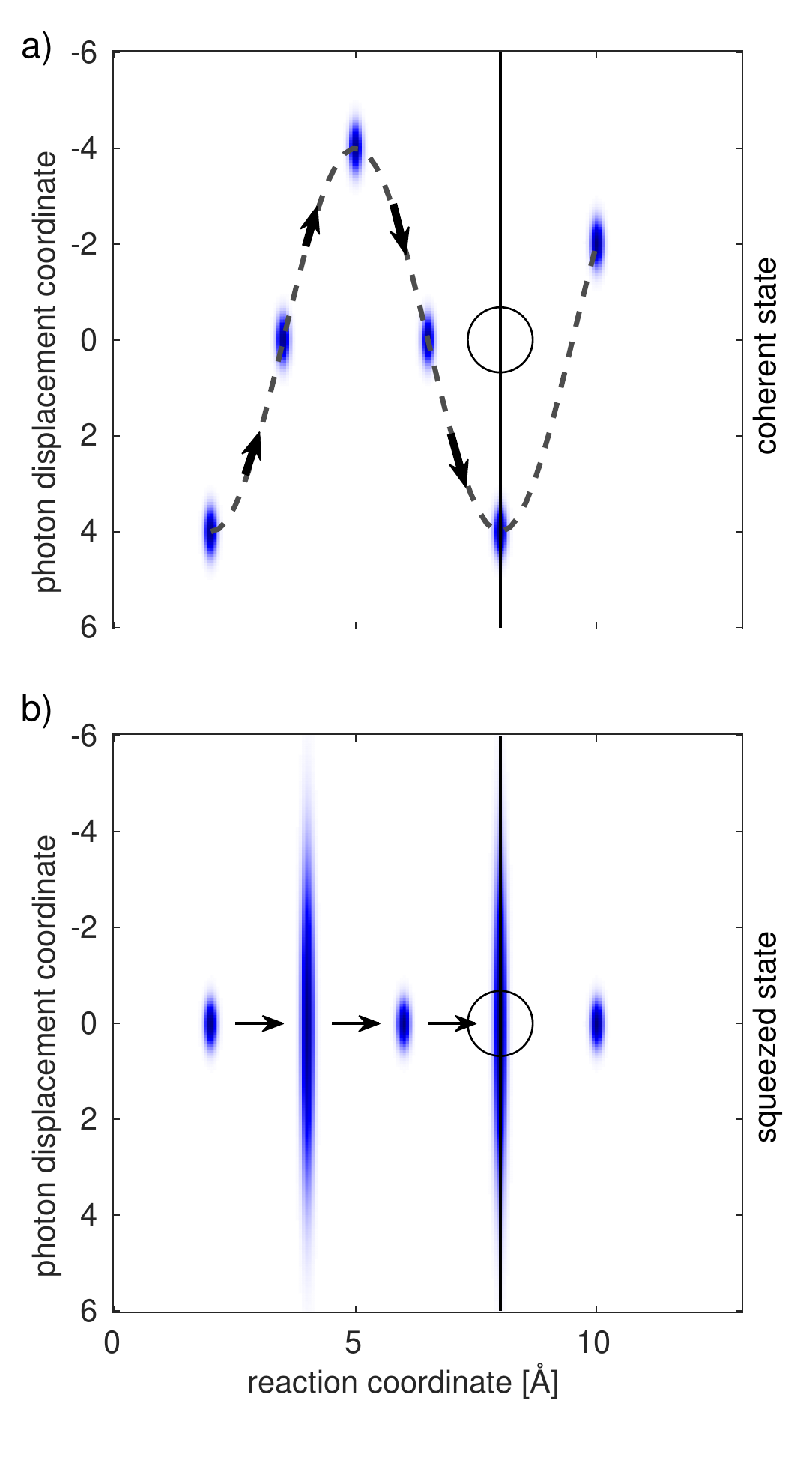}
\caption{Illustration of the interaction of a molecule with a coherent state (a) and squeezed vacuum state (b)
in a wave packet picture. The nuclear wave packet follows the gradient on the potential energy curve from the Franck-Condon point ($\approx 1.6$\,\AA) towards the avoided crossing (black circle at $\approx 8$\,\AA).
}
\label{fig:scheme}
\end{figure}
Figure \ref{fig:scheme}(b) illustrates the basic principle for a squeezed vacuum state
in the joint nuclear-photonic subspace. The initial state is a product state made up of
the vibrational ground state located at an internuclear separation of 1.6\,\AA\ and a squeezed vacuum state
centered around a photon displacement coordinate of 0. As the nuclear wave packet
in the excited electronic state follows the gradient towards the avoided crossing at 8.1\,\AA\ (which also the point of resonance),
the photon wave packet executes a breathing motion in $x$. By controlling
the initial phase of the squeezed state one can control the phase of the breathing motion
and thus control the strength of the interaction at the point in time when the
molecule reaches the point of resonance. Since the interaction is proportional to $\hat x$ the width of the photonic
wave packet at an instant in time will determine the effective strength of the interaction, when the molecule
reaches the point of resonance.
In Fig.\ \ref{fig:scheme}(a) we illustrate the same control principle but
with a coherent state. Here we can choose the initial momentum and displacement,
which is equivalent of choosing phase and amplitude of a classical laser field. The displacement
of the photon mode, when the molecule reaches the resonance point, will decide the strength
of the interaction. Combining a coherent state and a squeezed state yields a coherent squeezed state
and we now have the squeezing phase and the phase of the coherent state as control parameters.

The squeezing motion and the motion of the coherent state depend on the frequency of the
light mode $\omega_c$. To effectively use their motion to control the molecular degrees of freedom
the frequency of the photon mode needs to be on the similar time scale than the nuclear time
evolution.

\section{Results and Discussion}
The initial state of the time evolution is a product state of the photon mode (see Eqs. \ref{eq:coh}, \ref{eq:sqvc}, or \ref{eq:sqcoh}), the vibrational ground state of LiF and the electronic state $\Sigma_2$.
This corresponds to an impulsive
excitation with an ultra-short laser pulse to trigger nuclear dynamics. The initial state of the photon mode, that enters the product state represents the control parameters. In the following we will use different initial states for the photon mode to demonstrate the influence on the branching of the nuclear wave packet at the avoided crossing in LiF.
The frequency of the cavity mode is chosen such that is in resonance with the molecule exactly at the avoided crossing. Note that in Eq.\ \ref{eq:Hkl} we have neglected the permanent dipole
moments. Since the frequency of the cavity mode is in the infra-red regime it would couple directly
to the vibrational motion through the permanent dipole moments. We leave the investigation of
this effect to future work and focus only on the interaction with the electronic transition dipole moments.
The control objective is the population in the electronic ground state $\Sigma_1$ after 200\,fs, which is compared to the field free case.
The most obvious choice as an initial state is a Fock state. This has been already demonstrated for NaI in previous work \cite{Markus1}.
Pure Fock states have the most resemblance with classical light in terms of interaction and dynamics,
which has been demonstrated in \cite{Vibok1}. In case of a two-level
system their population dynamics are identical (a demonstration is given in Fig.\ S5 in the supplementary material). Single Fock states
do only offer the photon number $n$ as a control parameter but lack any form of phase control.
Consequently, Fock states are not considered here for control purposes.

\begin{figure}
\includegraphics[scale=0.5]{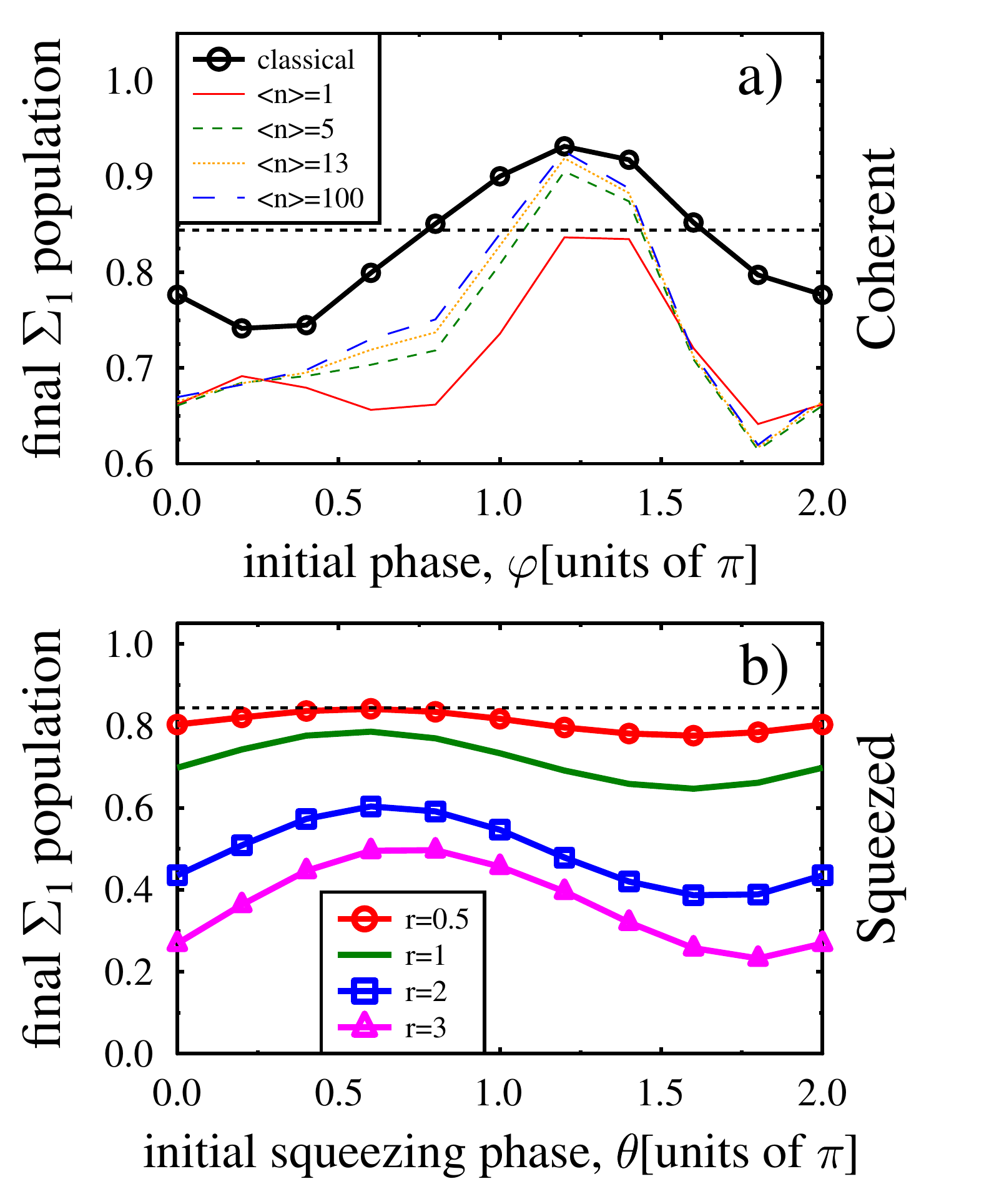}
\caption{Final $\Sigma_1$ state populations for (a) coherent and (b) squeezed vacuum initial states as a function of the initial phase and initial squeezing phase, respectively. In case of coherent states (a), several initial average photon numbers are considered. For comparison, the classical field description results are presented by the black line where the electric field amplitude is determined from $E_c=\chi \omega_c x_{max}$. In case of the coherent states the coupling strength parameter is scaled according to $\chi/\sqrt{\ev n}$. For the squeezed vacuum initial states (b), four different squeezing parameters are considered. In both panels $\omega_c$=0.037\,eV and $\chi$=0.01 are applied. The horizontal dashed lines show the field-free final populations in both panels.}\label{fig:squ_coh}
\end{figure}

\subsection{Coherent states and comparison with the classical state}
First, we compare different coherent states with each other, and its classical counter parts.
Coherent states are thought of as a close resemblance to classical coherent light,
since their time-dependent electric field expectation value
yields the classical electric field (see Eq.\ \ref{eq:Ecx} in the appendix).
However, the dynamics of the system only converges to a classical behaviour in the limit of
large photon numbers
(a Fock state within the Jaynes-Cummings model resembles the dynamics already for small photon numbers).
In the regime of small photon numbers the back-action of the molecule onto the field mode will cause a significant perturbation of the coherent state.
The initial state of the photon mode $\Psi_c$ is now given by Eqs.\ \ref{eq:coh}-\ref{eq:evp}.

In Fig.\ \ref{fig:squ_coh}(a) the results for coherent states with $\ev n=(1,5,13,100)$ are shown
(red, green, yellow, and blue curve respectively) alongside with the result for a classical field (black curve).
The field free case is denoted by the dashed line.
Here we use the coherent state phase $\varphi$ and the classical field phase $\phi$ as a control parameter.
Their coupling strengths are chosen such that the matrix elements of the light-matter
coupling are comparable in magnitude.
A clear variation of the final population ($t_{final}=200$\,fs) with respect to the phase can be observed.
The coherent states show a phase dependent modulation depth of 0.2 for
the single photon ($\ev n=1$) and converges to 0.3 for large photon numbers ($\ev n=100$).
The comparison with the classical field shows a comparable phase dependent modulation depth of 0.2 and it differs in the total suppression of the final population.
Note that control with a classical field or a coherent state enables
suppression as well as enhancement of the final population.

 \begin{table*}
 \footnotesize
 \caption{Relation between the initial photon number ($\ev{n}$), the minimum and maximum change in photon number ($\Delta \ev{n}$), as well as the minimum and maximum ground state populations (P) after the reaction has occurred in case of both the coherent and squeezed lights. For the squeezed states the r squeezing parameters are also shown. The data used to create this table can be found in Figs.\ S1-S3 in the supplementary material.}\label{tab:rnDeltan}
 \begin{tabular}{cccccccccc}\hline\hline
 Coherent light\\\hline
 $\alpha $ & $\ev{n}$ \quad & min. $\Delta \ev{n}$& $\varphi_{min}$ [$\pi$]& max. $\Delta \ev{n}$  & $\varphi_{max}$ [$\pi$] & $P_{min}$ & $\varphi_{min}$ [$\pi$] & $P_{max}$ & $\varphi_{max}$ [$\pi$]\\\hline
    1     & 1  &   -0.2  & 0.2 & 1   & 1.4 & 0.641 & 1.8 & 0.837 & 1.2\\
    $\sqrt{5}$    & 5  &   -0.5   & 0.2 & 0.7 & 1.4 & 0.661 & 1.8 & 0.906 & 1.2\\
    $\sqrt{13}$    & 13  &  -0.6   & 0.2 & 0.7 & 1.4 & 0.617 & 1.8 & 0.919 & 1.2\\
    10    & 100  & -0.6   & 0.2 & 0.7 & 1.4 & 0.620 & 1.8 & 0.926 & 1.2\\\hline\hline
Squeezed light\\\hline
 $r$ & $\ev{n}$ \quad & min. $\Delta \ev{n}$& $\theta_{min}$ [$\pi$] &max. $\Delta \ev{n}$  & $\theta_{max}$ [$\pi$]& $P_{min}$ & $\theta_{min}$ [$\pi$]& $P_{max}$ & $\theta_{max}$ [$\pi$]\\\hline
 0.5    & 0.3    & 0.2  & 0 & 0.4 & 1 & 0.776 & 1.6 & 0.842 & 0.6 \\
 1.0    & 1.4    & 0.1  & 0 & 0.8 & 1 & 0.647 & 1.6 & 0.786 & 0.6 \\
 2.0    & 13.2   & -0.4 & 0 & 3.2 & 1 & 0.387 & 1.6 & 0.604 & 0.6 \\
 3.0    & 100.4  & -2   & 0 & 7   & 1 & 0.233 & 1.8 & 0.497 & 0.8 \\\hline\hline
 \end{tabular}
 \end{table*}

\subsection{Squeezed Vacuum State}
 Next, we compare squeezed states with different squeezing parameters against each other. The initial state of the cavity mode is given by Eqs.\ \ref{eq:sqvc}-\ref{eq:Deltax}.
 This is a purely quantum mechanical state of light,
 which can not be represented by classical light.
 In Fig.\ \ref{fig:squ_coh}(b) the population in the $\Sigma_1$ state at the
 final time $t_{final}$ is plotted against the squeezing phase for different values of the squeezing parameter $r$ and a constant value for the coupling strength.
 The black dashed line in Fig.\ \ref{fig:squ_coh}(b) indicates the result
 of the photo-reaction without the influence of a cavity mode.
 For all values of $r$ we see a clear influence of $\theta$ on the final population.
 The result is a sinusoidal modulation with respect to the squeezing phase.
 The modulation depth increases with an increase of the squeezing parameter (values in table\ \ref{tab:rnDeltan}), ranging from a difference of 0.066 in the final $\Sigma_1$ population to
 0.26, for $r=0.5$ and $r=3$ respectively.
 Note that with an increase of $r$ the photon number $\ev{n}$ of the cavity also increases,
 leading to a stronger interaction (see table\ \ref{tab:rnDeltan}). This results in an
 increasingly suppressed dissociation, which may be explained by the increased separation of the dressed states leading to a decreased population exchange \cite{Kowalewski16jcp,Feist1}. For example for $r=3$ the approximate Rabi splitting is already 0.6\,eV.
 For all values of $r$ investigated here the final population is always suppressed
 compared to the field free case.

\subsection{Squeezed-Coherent states}
\begin{figure*}
\includegraphics[scale=1.1]{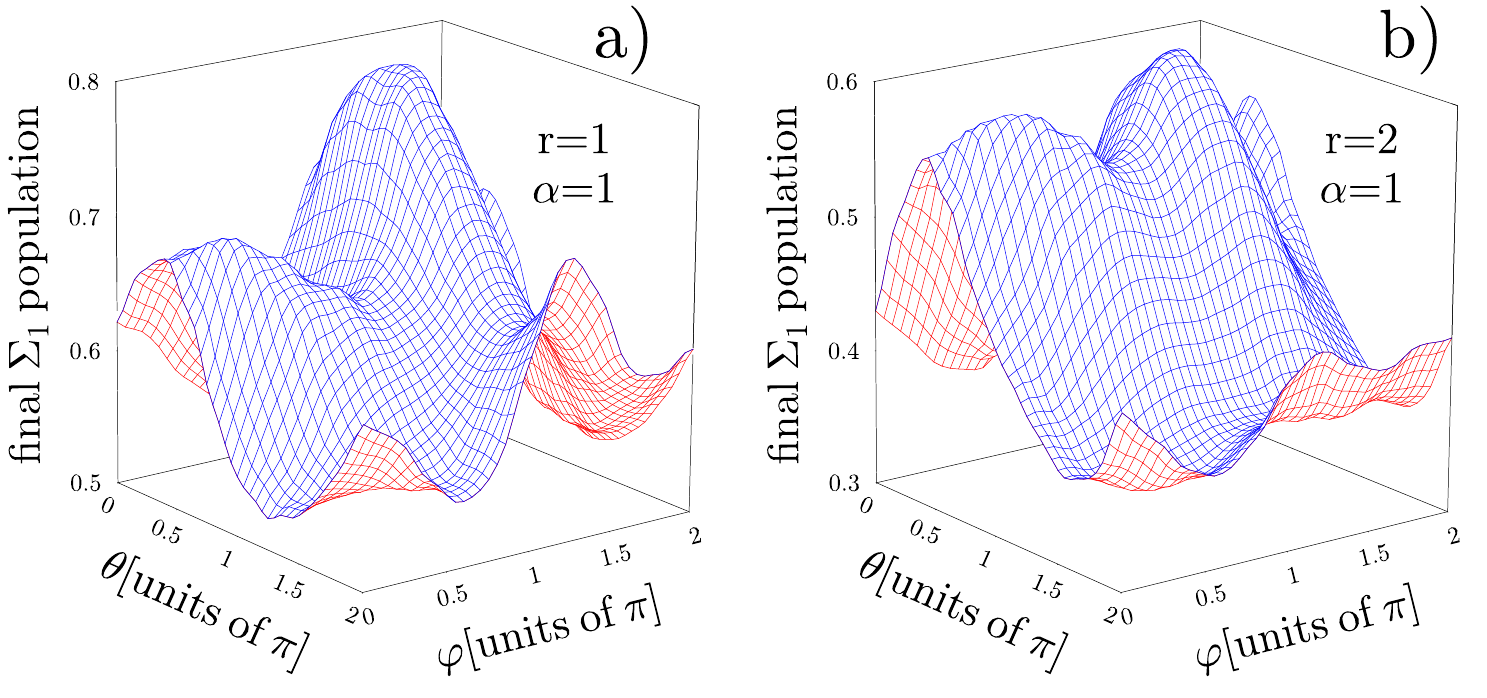}
\caption{Final $\Sigma_1$ state populations calculated as a function of the $\varphi$ initial phase and $\theta$ initial squeezing phase, using squeezed-coherent initial states.
The applied parameters are $|\alpha|$=1, r=1 (a) and $|\alpha|$=1, r=2 (b).
In both panels the coupling strength and transition frequency are $\chi$=0.01 and $\omega_c$=0.037 eV, respectively.
\label{fig:thetafi}}
\end{figure*}

We now discuss control via squeezed-coherent states. The initial state of the cavity mode can then be described by Eq.\ \ref{eq:sqcoh}.
Assuming that the displacement $|\alpha|$ and the squeezing parameter $r$ is kept constant we now have two phase variables that can be used to
control the final population: the phase space angle $\varphi$ of the coherent state and the squeezing phase $\theta$. In Fig.\ \ref{fig:thetafi} the final populations
are shown in dependence of $\theta$ and $\varphi$ for a coherent state displacement corresponding to $|\alpha|$=1 and two different squeezing parameters ($r=1$ and $r=2$).
Both control surfaces show clear local minima and maxima in the final $\Sigma_1$ population.
The control surface in Fig.\ \ref{fig:thetafi}(a) for $r=1$ varies from a final
population of 0.5 to 0.8, which is a larger variation than using  only a squeezed
state (Fig.\ \ref{fig:squ_coh}(b), green curve) or only a coherent state (Fig.\ \ref{fig:thetafi}(a)).
Increasing the squeezing parameter to $r=2$ in Fig.\ \ref{fig:thetafi}(b) results in
a stronger suppression of the $\Sigma_1$ population and the final
population now ranges from 0.3 to 0.6.
Both investigated cases allow only for suppression final population
(compared to field free $\approx 0.84$).
This trend may be explained by the trend that quantum light is suppressing the dissociation with increasing intensity. This also consistent with the blue curve from Fig.\ \ref{fig:squ_coh}(b) ($r=2$). The modulation depth (from global minima to global maxima)
is $\approx0.28$ in both cases. A noteworthy difference between Fig.\ \ref{fig:thetafi}(a) and
Fig.\ \ref{fig:thetafi}(b) is difference in the two local maxima at $\theta \approx 0.5\pi$ and the local minima at $\theta \approx 1.5\pi$:
for $r=1$ they differ by $\approx 0.1$, while for $r=2$ they are almost equal.

\subsection{Discussion}
We have investigated different quantum states of light with respect to their capability of modifying the dissociation behaviour at the avoided crossing in LiF and compared it to the control with classical single mode field.
Given that the frequency, polarization of the field, and the magnitude of the interaction are fixed, the only control parameter that the classical light field provides is the
carrier phase. The closest resemblance to this scenario
is a coherent state, which offers the phase $\varphi$ as a comparable parameter.
However, even if we fix the effective strength
of the interaction term by keeping $\chi \sqrt{n+1}$ constant, varying the photon number $n$ leads to different results.
This effect can be attributed to the fact the molecule can modify the photon mode.
A classical description corresponds to coherent state with a large photon number,
such that the exchange of a few photons does not affect the photonic wave packet.
The pictorial representation of the control principle in Fig.\ \ref{fig:scheme} is based on the idea that we can
control the shape of the wave packet in the photon displacement mode, which in turn controls the magnitude
of the interaction, when the molecule reaches the avoided crossing.
The investigated states, namely the coherent
states and the vacuum squeezed states are characterized by a sinusoidal time evolution of the photon displacement
and a sinusoidal time varying width of the photonic wave packet. This behavior is retrieved in the modulation
of the $\Sigma_1$ population for the coherent state phase and the squeezing phase.
The analogy in the classical picture is given by the instantaneous value of the electric field
when the molecule reaches the avoided crossing.
In the quantum description of light there is now more than one parameter to steer this effect.
Comparing the final populations of the squeezed states ($r=2$, Fig.\ \ref{fig:squ_coh}(b)) and the coherent states for a similar photon number ($\ev n$, Fig.\ \ref{fig:squ_coh}(a)), one finds a similar variation in the $\Sigma_1$ population of $\approx 0.2$.
The squeezed-coherent state shows a higher controllability with a difference in the $\Sigma_1$ population of $\approx 0.28$. Comparing this feature to Fig.\ \ref{fig:squ_coh} it allows
for a higher degree of control over the variation in final population in $\Sigma_1$ than either the squeezed vacuum or the coherent state alone.
However, classical light and coherent states are found to allow
for suppression or enhancement of the $\Sigma_1$ population while for squeezed vacuum states
and squeezed coherent states only a suppression of the $\Sigma_1$ population was observed.

\section{Conclusions and outlook}
We could show that quantum light in a cavity may be used to control
nonadiabatic dynamics in LiF. The squeezed state phase and/or the
coherent states can be used to alter the dissociation rate via the
$\Sigma_1$ state.
The presented control scheme relies on a fixed phase between an external pump-pulse, triggering the nuclear dynamics, and the initial state of the photon mode.
How the initial state of the cavity could be prepared in an experiment is an open question.
For the generation of squeezed-coherent states non-linear optical processes such as
optical parametric oscillators \cite{Hetet06jpb} or parametric down conversion \cite{Ast13oe} may be used.
The externally generated, non-classical, light then needs to be transferred to the cavity mode
containing the molecule.

Future investigations should involve a multi-mode description.
This will allow for a comparison with classical shaped laser pulses.
The relative phases between the field modes can be expected to become important
extending the control scheme significantly. A single light mode can only use
the carrier of the light wave to modulate the interaction strength
in time domain. However, a multi-mode scheme would recover the behavior of laser pulses, which are essentially multi-mode classical light fields \cite{Gruebel01cp}.
This allows for control of time scales much smaller than the oscillation
period of the carrier frequency.

Moreover, one may envision to extend the presented principle to arbitrary quantum light states.
Optimal control theory would then optimize an initial quantum state of the cavity modes rather than
the classical phase-amplitude shape of a light field.
Moreover, an interesting field of study maybe the application of the control scheme to collectively coupled ensembles \cite{Feist4,Vendrell18prl} of molecules. The collective enhancement may be controlled
by means of the quantum state of the cavity mode.

\begin{acknowledgments}
This research was supported by the EU-funded Hungarian Grant No. EFOP-3.6.2-16-2017-00005. The authors
are grateful to NKFIH for support (Grant No. K128396).
M.K. acknowledges support from the Swedish Research Council (Grant No. 2018-05346).
\end{acknowledgments}

\bibliography{QuantumLightControl.bib}

\begin{thebibliography}{57}%
\makeatletter
\providecommand \@ifxundefined [1]{%
 \@ifx{#1\undefined}
}%
\providecommand \@ifnum [1]{%
 \ifnum #1\expandafter \@firstoftwo
 \else \expandafter \@secondoftwo
 \fi
}%
\providecommand \@ifx [1]{%
 \ifx #1\expandafter \@firstoftwo
 \else \expandafter \@secondoftwo
 \fi
}%
\providecommand \natexlab [1]{#1}%
\providecommand \enquote  [1]{``#1''}%
\providecommand \bibnamefont  [1]{#1}%
\providecommand \bibfnamefont [1]{#1}%
\providecommand \citenamefont [1]{#1}%
\providecommand \href@noop [0]{\@secondoftwo}%
\providecommand \href [0]{\begingroup \@sanitize@url \@href}%
\providecommand \@href[1]{\@@startlink{#1}\@@href}%
\providecommand \@@href[1]{\endgroup#1\@@endlink}%
\providecommand \@sanitize@url [0]{\catcode `\\12\catcode `\$12\catcode
  `\&12\catcode `\#12\catcode `\^12\catcode `\_12\catcode `\%12\relax}%
\providecommand \@@startlink[1]{}%
\providecommand \@@endlink[0]{}%
\providecommand \url  [0]{\begingroup\@sanitize@url \@url }%
\providecommand \@url [1]{\endgroup\@href {#1}{\urlprefix }}%
\providecommand \urlprefix  [0]{URL }%
\providecommand \Eprint [0]{\href }%
\providecommand \doibase [0]{http://dx.doi.org/}%
\providecommand \selectlanguage [0]{\@gobble}%
\providecommand \bibinfo  [0]{\@secondoftwo}%
\providecommand \bibfield  [0]{\@secondoftwo}%
\providecommand \translation [1]{[#1]}%
\providecommand \BibitemOpen [0]{}%
\providecommand \bibitemStop [0]{}%
\providecommand \bibitemNoStop [0]{.\EOS\space}%
\providecommand \EOS [0]{\spacefactor3000\relax}%
\providecommand \BibitemShut  [1]{\csname bibitem#1\endcsname}%
\let\auto@bib@innerbib\@empty
\bibitem [{\citenamefont {Warren}\ \emph {et~al.}(1993)\citenamefont {Warren},
  \citenamefont {Rabitz},\ and\ \citenamefont {Dahleh}}]{Warren93}%
  \BibitemOpen
  \bibfield  {author} {\bibinfo {author} {\bibfnamefont {W.~S.}\ \bibnamefont
  {Warren}}, \bibinfo {author} {\bibfnamefont {H.}~\bibnamefont {Rabitz}}, \
  and\ \bibinfo {author} {\bibfnamefont {M.}~\bibnamefont {Dahleh}},\ }\href
  {\doibase 10.1126/science.259.5101.1581} {\bibfield  {journal} {\bibinfo
  {journal} {Science}\ }\textbf {\bibinfo {volume} {259}},\ \bibinfo {pages}
  {1581} (\bibinfo {year} {1993})}\BibitemShut {NoStop}%
\bibitem [{\citenamefont {Bartanaa}\ \emph {et~al.}(2001)\citenamefont
  {Bartanaa}, \citenamefont {Kosloff},\ and\ \citenamefont
  {Tannor}}]{Bartanaa01cp}%
  \BibitemOpen
  \bibfield  {author} {\bibinfo {author} {\bibfnamefont {A.}~\bibnamefont
  {Bartanaa}}, \bibinfo {author} {\bibfnamefont {R.}~\bibnamefont {Kosloff}}, \
  and\ \bibinfo {author} {\bibfnamefont {D.~J.}\ \bibnamefont {Tannor}},\
  }\href {\doibase 10.1016/S0301-0104(01)00266-X} {\bibfield  {journal}
  {\bibinfo  {journal} {Chem. Phys.}\ }\textbf {\bibinfo {volume} {267}},\
  \bibinfo {pages} {195} (\bibinfo {year} {2001})}\BibitemShut {NoStop}%
\bibitem [{\citenamefont {von~den Hoff}\ \emph {et~al.}(2012)\citenamefont
  {von~den Hoff}, \citenamefont {Thallmair}, \citenamefont {Kowalewski},
  \citenamefont {Siemering},\ and\ \citenamefont
  {de~Vivie-Riedle}}]{vdHoff12pccp}%
  \BibitemOpen
  \bibfield  {author} {\bibinfo {author} {\bibfnamefont {P.}~\bibnamefont
  {von~den Hoff}}, \bibinfo {author} {\bibfnamefont {S.}~\bibnamefont
  {Thallmair}}, \bibinfo {author} {\bibfnamefont {M.}~\bibnamefont
  {Kowalewski}}, \bibinfo {author} {\bibfnamefont {R.}~\bibnamefont
  {Siemering}}, \ and\ \bibinfo {author} {\bibfnamefont {R.}~\bibnamefont
  {de~Vivie-Riedle}},\ }\href {\doibase 10.1039/c2cp41838j} {\bibfield
  {journal} {\bibinfo  {journal} {Phys. Chem. Chem. Phys.}\ }\textbf {\bibinfo
  {volume} {14}},\ \bibinfo {pages} {14460+} (\bibinfo {year}
  {2012})}\BibitemShut {NoStop}%
\bibitem [{\citenamefont {Koch}\ and\ \citenamefont
  {Shapiro}(2012)}]{Koch12cr}%
  \BibitemOpen
  \bibfield  {author} {\bibinfo {author} {\bibfnamefont {C.~P.}\ \bibnamefont
  {Koch}}\ and\ \bibinfo {author} {\bibfnamefont {M.}~\bibnamefont {Shapiro}},\
  }\href {\doibase 10.1021/cr2003882} {\bibfield  {journal} {\bibinfo
  {journal} {Chemical Reviews}\ }\textbf {\bibinfo {volume} {112}},\ \bibinfo
  {pages} {4928} (\bibinfo {year} {2012})}\BibitemShut {NoStop}%
\bibitem [{\citenamefont {Dantus}\ and\ \citenamefont
  {Lozovoy}(2004)}]{Dantus04cr}%
  \BibitemOpen
  \bibfield  {author} {\bibinfo {author} {\bibfnamefont {M.}~\bibnamefont
  {Dantus}}\ and\ \bibinfo {author} {\bibfnamefont {V.~V.}\ \bibnamefont
  {Lozovoy}},\ }\href {\doibase 10.1021/cr020668r} {\bibfield  {journal}
  {\bibinfo  {journal} {Chemical Reviews}\ }\textbf {\bibinfo {volume} {104}},\
  \bibinfo {pages} {1813} (\bibinfo {year} {2004})}\BibitemShut {NoStop}%
\bibitem [{\citenamefont {Shapiro}\ and\ \citenamefont
  {Brumer}(2012)}]{Shapiro}%
  \BibitemOpen
  \bibfield  {author} {\bibinfo {author} {\bibfnamefont {M.}~\bibnamefont
  {Shapiro}}\ and\ \bibinfo {author} {\bibfnamefont {P.}~\bibnamefont
  {Brumer}},\ }\href {\doibase 10.1002/9783527639700} {\emph {\bibinfo {title}
  {Quantum Control of Molecular Processes}}},\ \bibinfo {edition} {2nd}\ ed.\
  (\bibinfo  {publisher} {Wiley‐VCH Verlag GmbH \& Co. KGaA},\ \bibinfo
  {year} {2012})\BibitemShut {NoStop}%
\bibitem [{\citenamefont {Brif}\ \emph {et~al.}(2010)\citenamefont {Brif},
  \citenamefont {Chakrabarti},\ and\ \citenamefont {Rabitz}}]{Brif10njp}%
  \BibitemOpen
  \bibfield  {author} {\bibinfo {author} {\bibfnamefont {C.}~\bibnamefont
  {Brif}}, \bibinfo {author} {\bibfnamefont {R.}~\bibnamefont {Chakrabarti}}, \
  and\ \bibinfo {author} {\bibfnamefont {H.}~\bibnamefont {Rabitz}},\ }\href
  {\doibase 10.1088/1367-2630/12/7/075008} {\bibfield  {journal} {\bibinfo
  {journal} {New J. Phys.}\ }\textbf {\bibinfo {volume} {12}},\ \bibinfo
  {pages} {075008} (\bibinfo {year} {2010})}\BibitemShut {NoStop}%
\bibitem [{\citenamefont {Brumer}\ and\ \citenamefont
  {Shapiro}(1986)}]{Brumer86cpl}%
  \BibitemOpen
  \bibfield  {author} {\bibinfo {author} {\bibfnamefont {P.}~\bibnamefont
  {Brumer}}\ and\ \bibinfo {author} {\bibfnamefont {M.}~\bibnamefont
  {Shapiro}},\ }\href {\doibase 10.1016/s0009-2614(86)80171-3} {\bibfield
  {journal} {\bibinfo  {journal} {Chem. Phys. Lett.}\ }\textbf {\bibinfo
  {volume} {126}},\ \bibinfo {pages} {541} (\bibinfo {year}
  {1986})}\BibitemShut {NoStop}%
\bibitem [{\citenamefont {Shapiro}\ \emph {et~al.}(1988)\citenamefont
  {Shapiro}, \citenamefont {Hepburn},\ and\ \citenamefont
  {Brumer}}]{Shapiro88cpl}%
  \BibitemOpen
  \bibfield  {author} {\bibinfo {author} {\bibfnamefont {M.}~\bibnamefont
  {Shapiro}}, \bibinfo {author} {\bibfnamefont {J.~W.}\ \bibnamefont
  {Hepburn}}, \ and\ \bibinfo {author} {\bibfnamefont {P.}~\bibnamefont
  {Brumer}},\ }\href {\doibase 10.1016/0009-2614(88)80362-2} {\bibfield
  {journal} {\bibinfo  {journal} {Chem. Phys. Lett.}\ }\textbf {\bibinfo
  {volume} {149}},\ \bibinfo {pages} {451} (\bibinfo {year}
  {1988})}\BibitemShut {NoStop}%
\bibitem [{\citenamefont {Tannor}\ \emph {et~al.}(1986)\citenamefont {Tannor},
  \citenamefont {Kosloff},\ and\ \citenamefont {Rice}}]{Tannor86jcp}%
  \BibitemOpen
  \bibfield  {author} {\bibinfo {author} {\bibfnamefont {D.~J.}\ \bibnamefont
  {Tannor}}, \bibinfo {author} {\bibfnamefont {R.}~\bibnamefont {Kosloff}}, \
  and\ \bibinfo {author} {\bibfnamefont {S.~A.}\ \bibnamefont {Rice}},\ }\href
  {\doibase 10.1063/1.451542} {\bibfield  {journal} {\bibinfo  {journal} {J.
  Chem. Phys.}\ }\textbf {\bibinfo {volume} {85}},\ \bibinfo {pages} {5805}
  (\bibinfo {year} {1986})}\BibitemShut {NoStop}%
\bibitem [{\citenamefont {Gordon}\ and\ \citenamefont
  {Rice}(1997)}]{Gordon97arpc}%
  \BibitemOpen
  \bibfield  {author} {\bibinfo {author} {\bibfnamefont {R.~J.}\ \bibnamefont
  {Gordon}}\ and\ \bibinfo {author} {\bibfnamefont {S.~A.}\ \bibnamefont
  {Rice}},\ }\href {\doibase 10.1146/annurev.physchem.48.1.601} {\bibfield
  {journal} {\bibinfo  {journal} {Annu. Rev. Phys. Chem.}\ }\textbf {\bibinfo
  {volume} {48}},\ \bibinfo {pages} {601} (\bibinfo {year} {1997})}\BibitemShut
  {NoStop}%
\bibitem [{\citenamefont {Aspelmeyer}\ \emph {et~al.}(2014)\citenamefont
  {Aspelmeyer}, \citenamefont {Kippenberg},\ and\ \citenamefont
  {Marquardt}}]{Aspelmeyer}%
  \BibitemOpen
  \bibfield  {author} {\bibinfo {author} {\bibfnamefont {M.}~\bibnamefont
  {Aspelmeyer}}, \bibinfo {author} {\bibfnamefont {T.~J.}\ \bibnamefont
  {Kippenberg}}, \ and\ \bibinfo {author} {\bibfnamefont {F.}~\bibnamefont
  {Marquardt}},\ }\href@noop {} {\bibfield  {journal} {\bibinfo  {journal}
  {Rev. Mod. Phys.}\ }\textbf {\bibinfo {volume} {86}},\ \bibinfo {pages}
  {1391} (\bibinfo {year} {2014})}\BibitemShut {NoStop}%
\bibitem [{\citenamefont {Hutchison}\ \emph {et~al.}(2012)\citenamefont
  {Hutchison}, \citenamefont {Schwartz}, \citenamefont {Genet}, \citenamefont
  {Devaux},\ and\ \citenamefont {Ebbesen}}]{Ebbesen1}%
  \BibitemOpen
  \bibfield  {author} {\bibinfo {author} {\bibfnamefont {J.~A.}\ \bibnamefont
  {Hutchison}}, \bibinfo {author} {\bibfnamefont {T.}~\bibnamefont {Schwartz}},
  \bibinfo {author} {\bibfnamefont {C.}~\bibnamefont {Genet}}, \bibinfo
  {author} {\bibfnamefont {E.}~\bibnamefont {Devaux}}, \ and\ \bibinfo {author}
  {\bibfnamefont {T.~W.}\ \bibnamefont {Ebbesen}},\ }\href {\doibase
  10.1002/anie.201107033} {\bibfield  {journal} {\bibinfo  {journal} {Angew.
  Chem. Int. Ed.}\ }\textbf {\bibinfo {volume} {51}},\ \bibinfo {pages} {1592}
  (\bibinfo {year} {2012})}\BibitemShut {NoStop}%
\bibitem [{\citenamefont {Schwartz}\ \emph {et~al.}(2013)\citenamefont
  {Schwartz}, \citenamefont {Hutchison}, \citenamefont {L{\'{e}}onard},
  \citenamefont {Genet}, \citenamefont {Haacke},\ and\ \citenamefont
  {Ebbesen}}]{Ebbesen2}%
  \BibitemOpen
  \bibfield  {author} {\bibinfo {author} {\bibfnamefont {T.}~\bibnamefont
  {Schwartz}}, \bibinfo {author} {\bibfnamefont {J.~A.}\ \bibnamefont
  {Hutchison}}, \bibinfo {author} {\bibfnamefont {J.}~\bibnamefont
  {L{\'{e}}onard}}, \bibinfo {author} {\bibfnamefont {C.}~\bibnamefont
  {Genet}}, \bibinfo {author} {\bibfnamefont {S.}~\bibnamefont {Haacke}}, \
  and\ \bibinfo {author} {\bibfnamefont {T.~W.}\ \bibnamefont {Ebbesen}},\
  }\href {\doibase 10.1002/cphc.201200734} {\bibfield  {journal} {\bibinfo
  {journal} {Chem. Phys. Chem.}\ }\textbf {\bibinfo {volume} {14}},\ \bibinfo
  {pages} {125} (\bibinfo {year} {2013})}\BibitemShut {NoStop}%
\bibitem [{\citenamefont {George}\ \emph {et~al.}(2015)\citenamefont {George},
  \citenamefont {Wang}, \citenamefont {Chervy}, \citenamefont
  {Canaguier-Durand}, \citenamefont {Schaeffer}, \citenamefont {Lehn},
  \citenamefont {Hutchison}, \citenamefont {Genet},\ and\ \citenamefont
  {Ebbesen}}]{Ebbesen3}%
  \BibitemOpen
  \bibfield  {author} {\bibinfo {author} {\bibfnamefont {J.}~\bibnamefont
  {George}}, \bibinfo {author} {\bibfnamefont {S.}~\bibnamefont {Wang}},
  \bibinfo {author} {\bibfnamefont {T.}~\bibnamefont {Chervy}}, \bibinfo
  {author} {\bibfnamefont {A.}~\bibnamefont {Canaguier-Durand}}, \bibinfo
  {author} {\bibfnamefont {G.}~\bibnamefont {Schaeffer}}, \bibinfo {author}
  {\bibfnamefont {J.-M.}\ \bibnamefont {Lehn}}, \bibinfo {author}
  {\bibfnamefont {J.~A.}\ \bibnamefont {Hutchison}}, \bibinfo {author}
  {\bibfnamefont {C.}~\bibnamefont {Genet}}, \ and\ \bibinfo {author}
  {\bibfnamefont {T.~W.}\ \bibnamefont {Ebbesen}},\ }\href {\doibase
  10.1039/C4FD00197D} {\bibfield  {journal} {\bibinfo  {journal} {Faraday
  Discuss.}\ }\textbf {\bibinfo {volume} {178}},\ \bibinfo {pages} {281}
  (\bibinfo {year} {2015})}\BibitemShut {NoStop}%
\bibitem [{\citenamefont {Zhong}\ \emph {et~al.}(2016)\citenamefont {Zhong},
  \citenamefont {Chervy}, \citenamefont {Wang}, \citenamefont {George},
  \citenamefont {Thomas}, \citenamefont {Hutchison}, \citenamefont {Devaux},
  \citenamefont {Genet},\ and\ \citenamefont {Ebbesen}}]{Ebbesen4}%
  \BibitemOpen
  \bibfield  {author} {\bibinfo {author} {\bibfnamefont {X.}~\bibnamefont
  {Zhong}}, \bibinfo {author} {\bibfnamefont {T.}~\bibnamefont {Chervy}},
  \bibinfo {author} {\bibfnamefont {S.}~\bibnamefont {Wang}}, \bibinfo {author}
  {\bibfnamefont {J.}~\bibnamefont {George}}, \bibinfo {author} {\bibfnamefont
  {A.}~\bibnamefont {Thomas}}, \bibinfo {author} {\bibfnamefont {J.~A.}\
  \bibnamefont {Hutchison}}, \bibinfo {author} {\bibfnamefont {E.}~\bibnamefont
  {Devaux}}, \bibinfo {author} {\bibfnamefont {C.}~\bibnamefont {Genet}}, \
  and\ \bibinfo {author} {\bibfnamefont {T.~W.}\ \bibnamefont {Ebbesen}},\
  }\href {\doibase 10.1002/anie.201600428} {\bibfield  {journal} {\bibinfo
  {journal} {Angew. Chem. Int. Ed.}\ }\textbf {\bibinfo {volume} {55}},\
  \bibinfo {pages} {6202} (\bibinfo {year} {2016})}\BibitemShut {NoStop}%
\bibitem [{\citenamefont {Ebbesen}(2016)}]{Ebbesen5}%
  \BibitemOpen
  \bibfield  {author} {\bibinfo {author} {\bibfnamefont {T.~W.}\ \bibnamefont
  {Ebbesen}},\ }\href {\doibase 10.1021/acs.accounts.6b00295} {\bibfield
  {journal} {\bibinfo  {journal} {Acc. Chem. Res.}\ }\textbf {\bibinfo {volume}
  {49}},\ \bibinfo {pages} {2403} (\bibinfo {year} {2016})}\BibitemShut
  {NoStop}%
\bibitem [{\citenamefont {Thomas}\ \emph {et~al.}(2016)\citenamefont {Thomas},
  \citenamefont {George}, \citenamefont {Shalabney}, \citenamefont {Dryzhakov},
  \citenamefont {Varma}, \citenamefont {Moran}, \citenamefont {Chervy},
  \citenamefont {Zhong}, \citenamefont {Devaux}, \citenamefont {Genet},
  \citenamefont {Hutchison},\ and\ \citenamefont {Ebbesen}}]{Ebbesen6}%
  \BibitemOpen
  \bibfield  {author} {\bibinfo {author} {\bibfnamefont {A.}~\bibnamefont
  {Thomas}}, \bibinfo {author} {\bibfnamefont {J.}~\bibnamefont {George}},
  \bibinfo {author} {\bibfnamefont {A.}~\bibnamefont {Shalabney}}, \bibinfo
  {author} {\bibfnamefont {M.}~\bibnamefont {Dryzhakov}}, \bibinfo {author}
  {\bibfnamefont {S.~J.}\ \bibnamefont {Varma}}, \bibinfo {author}
  {\bibfnamefont {J.}~\bibnamefont {Moran}}, \bibinfo {author} {\bibfnamefont
  {T.}~\bibnamefont {Chervy}}, \bibinfo {author} {\bibfnamefont
  {X.}~\bibnamefont {Zhong}}, \bibinfo {author} {\bibfnamefont
  {E.}~\bibnamefont {Devaux}}, \bibinfo {author} {\bibfnamefont
  {C.}~\bibnamefont {Genet}}, \bibinfo {author} {\bibfnamefont {J.~A.}\
  \bibnamefont {Hutchison}}, \ and\ \bibinfo {author} {\bibfnamefont {T.~W.}\
  \bibnamefont {Ebbesen}},\ }\href {\doibase 10.1002/anie.201605504} {\bibfield
   {journal} {\bibinfo  {journal} {Angew. Chem. Int. Ed.}\ }\textbf {\bibinfo
  {volume} {55}},\ \bibinfo {pages} {11462} (\bibinfo {year}
  {2016})}\BibitemShut {NoStop}%
\bibitem [{\citenamefont {Galego}\ \emph {et~al.}(2015)\citenamefont {Galego},
  \citenamefont {Garcia-Vidal},\ and\ \citenamefont {Feist}}]{Feist1}%
  \BibitemOpen
  \bibfield  {author} {\bibinfo {author} {\bibfnamefont {J.}~\bibnamefont
  {Galego}}, \bibinfo {author} {\bibfnamefont {F.~J.}\ \bibnamefont
  {Garcia-Vidal}}, \ and\ \bibinfo {author} {\bibfnamefont {J.}~\bibnamefont
  {Feist}},\ }\href {\doibase 10.1103/PhysRevX.5.041022} {\bibfield  {journal}
  {\bibinfo  {journal} {Phys. Rev. X}\ }\textbf {\bibinfo {volume} {5}},\
  \bibinfo {pages} {1} (\bibinfo {year} {2015})}\BibitemShut {NoStop}%
\bibitem [{\citenamefont {Galego}\ \emph {et~al.}(2016)\citenamefont {Galego},
  \citenamefont {Garcia-Vidal},\ and\ \citenamefont {Feist}}]{Feist2}%
  \BibitemOpen
  \bibfield  {author} {\bibinfo {author} {\bibfnamefont {J.}~\bibnamefont
  {Galego}}, \bibinfo {author} {\bibfnamefont {F.~J.}\ \bibnamefont
  {Garcia-Vidal}}, \ and\ \bibinfo {author} {\bibfnamefont {J.}~\bibnamefont
  {Feist}},\ }\href {\doibase 10.1038/ncomms13841} {\bibfield  {journal}
  {\bibinfo  {journal} {Nat. Commun.}\ }\textbf {\bibinfo {volume} {7}},\
  \bibinfo {pages} {1} (\bibinfo {year} {2016})}\BibitemShut {NoStop}%
\bibitem [{\citenamefont {Galego}\ \emph {et~al.}(2017)\citenamefont {Galego},
  \citenamefont {Garcia-Vidal},\ and\ \citenamefont {Feist}}]{Feist3}%
  \BibitemOpen
  \bibfield  {author} {\bibinfo {author} {\bibfnamefont {J.}~\bibnamefont
  {Galego}}, \bibinfo {author} {\bibfnamefont {F.~J.}\ \bibnamefont
  {Garcia-Vidal}}, \ and\ \bibinfo {author} {\bibfnamefont {J.}~\bibnamefont
  {Feist}},\ }\href {\doibase 10.1103/PhysRevLett.119.136001} {\bibfield
  {journal} {\bibinfo  {journal} {Phys. Rev. Lett.}\ }\textbf {\bibinfo
  {volume} {119}},\ \bibinfo {pages} {136001} (\bibinfo {year}
  {2017})}\BibitemShut {NoStop}%
\bibitem [{\citenamefont {Feist}\ \emph {et~al.}(2018)\citenamefont {Feist},
  \citenamefont {Galego},\ and\ \citenamefont {Garcia-Vidal}}]{Feist4}%
  \BibitemOpen
  \bibfield  {author} {\bibinfo {author} {\bibfnamefont {J.}~\bibnamefont
  {Feist}}, \bibinfo {author} {\bibfnamefont {J.}~\bibnamefont {Galego}}, \
  and\ \bibinfo {author} {\bibfnamefont {F.~J.}\ \bibnamefont {Garcia-Vidal}},\
  }\href {\doibase 10.1021/acsphotonics.7b00680} {\bibfield  {journal}
  {\bibinfo  {journal} {ACS Photonics}\ }\textbf {\bibinfo {volume} {5}},\
  \bibinfo {pages} {205} (\bibinfo {year} {2018})}\BibitemShut {NoStop}%
\bibitem [{\citenamefont {Luk}\ \emph {et~al.}(2017)\citenamefont {Luk},
  \citenamefont {Feist}, \citenamefont {Toppari},\ and\ \citenamefont
  {Groenhof}}]{Feist5}%
  \BibitemOpen
  \bibfield  {author} {\bibinfo {author} {\bibfnamefont {H.~L.}\ \bibnamefont
  {Luk}}, \bibinfo {author} {\bibfnamefont {J.}~\bibnamefont {Feist}}, \bibinfo
  {author} {\bibfnamefont {J.~J.}\ \bibnamefont {Toppari}}, \ and\ \bibinfo
  {author} {\bibfnamefont {G.}~\bibnamefont {Groenhof}},\ }\href {\doibase
  10.1021/acs.jctc.7b00388} {\bibfield  {journal} {\bibinfo  {journal} {J.
  Chem. Theory Comput.}\ }\textbf {\bibinfo {volume} {13}},\ \bibinfo {pages}
  {4324} (\bibinfo {year} {2017})}\BibitemShut {NoStop}%
\bibitem [{\citenamefont {Herrera}\ and\ \citenamefont
  {Spano}(2016)}]{Herrera1}%
  \BibitemOpen
  \bibfield  {author} {\bibinfo {author} {\bibfnamefont {F.}~\bibnamefont
  {Herrera}}\ and\ \bibinfo {author} {\bibfnamefont {F.~C.}\ \bibnamefont
  {Spano}},\ }\href {\doibase 10.1103/PhysRevLett.116.238301} {\bibfield
  {journal} {\bibinfo  {journal} {Phys. Rev. Lett.}\ }\textbf {\bibinfo
  {volume} {116}},\ \bibinfo {pages} {238301} (\bibinfo {year}
  {2016})}\BibitemShut {NoStop}%
\bibitem [{\citenamefont {Herrera}\ and\ \citenamefont
  {Spano}(2017)}]{Herrera2}%
  \BibitemOpen
  \bibfield  {author} {\bibinfo {author} {\bibfnamefont {F.}~\bibnamefont
  {Herrera}}\ and\ \bibinfo {author} {\bibfnamefont {F.~C.}\ \bibnamefont
  {Spano}},\ }\href {\doibase 10.1103/PhysRevLett.118.223601} {\bibfield
  {journal} {\bibinfo  {journal} {Phys. Rev. Lett.}\ }\textbf {\bibinfo
  {volume} {118}},\ \bibinfo {pages} {223601} (\bibinfo {year}
  {2017})}\BibitemShut {NoStop}%
\bibitem [{\citenamefont {Herrera}\ and\ \citenamefont
  {Spano}(2018)}]{Herrera3}%
  \BibitemOpen
  \bibfield  {author} {\bibinfo {author} {\bibfnamefont {F.}~\bibnamefont
  {Herrera}}\ and\ \bibinfo {author} {\bibfnamefont {F.~C.}\ \bibnamefont
  {Spano}},\ }\href {\doibase 10.1021/acsphotonics.7b00728} {\bibfield
  {journal} {\bibinfo  {journal} {ACS Photonics}\ }\textbf {\bibinfo {volume}
  {5}},\ \bibinfo {pages} {65} (\bibinfo {year} {2018})}\BibitemShut {NoStop}%
\bibitem [{\citenamefont {Ribeiro}\ \emph {et~al.}(2018)\citenamefont
  {Ribeiro}, \citenamefont {Mart\'inez-Mart\'inez}, \citenamefont {Du},
  \citenamefont {Campos-Gonzalez-Angulo},\ and\ \citenamefont
  {Yuen-Zhou}}]{Joel1}%
  \BibitemOpen
  \bibfield  {author} {\bibinfo {author} {\bibfnamefont {R.~F.}\ \bibnamefont
  {Ribeiro}}, \bibinfo {author} {\bibfnamefont {L.~A.}\ \bibnamefont
  {Mart\'inez-Mart\'inez}}, \bibinfo {author} {\bibfnamefont {M.}~\bibnamefont
  {Du}}, \bibinfo {author} {\bibfnamefont {J.}~\bibnamefont
  {Campos-Gonzalez-Angulo}}, \ and\ \bibinfo {author} {\bibfnamefont
  {J.}~\bibnamefont {Yuen-Zhou}},\ }\href {\doibase 10.1039/C8SC01043A}
  {\bibfield  {journal} {\bibinfo  {journal} {Chem. Sci.}\ }\textbf {\bibinfo
  {volume} {9}},\ \bibinfo {pages} {6325} (\bibinfo {year} {2018})}\BibitemShut
  {NoStop}%
\bibitem [{\citenamefont {Mart\'inez-Mart\'inez}\ \emph
  {et~al.}(2018)\citenamefont {Mart\'inez-Mart\'inez}, \citenamefont {Ribeiro},
  \citenamefont {Campos-Gonz\'alez-Angulo},\ and\ \citenamefont
  {Yuen-Zhou}}]{Joel2}%
  \BibitemOpen
  \bibfield  {author} {\bibinfo {author} {\bibfnamefont {L.~A.}\ \bibnamefont
  {Mart\'inez-Mart\'inez}}, \bibinfo {author} {\bibfnamefont {R.~F.}\
  \bibnamefont {Ribeiro}}, \bibinfo {author} {\bibfnamefont {J.}~\bibnamefont
  {Campos-Gonz\'alez-Angulo}}, \ and\ \bibinfo {author} {\bibfnamefont
  {J.}~\bibnamefont {Yuen-Zhou}},\ }\href {\doibase
  10.1021/acsphotonics.7b00610} {\bibfield  {journal} {\bibinfo  {journal} {ACS
  Photonics}\ }\textbf {\bibinfo {volume} {5}},\ \bibinfo {pages} {167}
  (\bibinfo {year} {2018})}\BibitemShut {NoStop}%
\bibitem [{\citenamefont {Yuen-Zhou}\ \emph {et~al.}(2018)\citenamefont
  {Yuen-Zhou}, \citenamefont {Saikin},\ and\ \citenamefont {Menon}}]{Joel3}%
  \BibitemOpen
  \bibfield  {author} {\bibinfo {author} {\bibfnamefont {J.}~\bibnamefont
  {Yuen-Zhou}}, \bibinfo {author} {\bibfnamefont {S.~K.}\ \bibnamefont
  {Saikin}}, \ and\ \bibinfo {author} {\bibfnamefont {V.~M.}\ \bibnamefont
  {Menon}},\ }\href {\doibase 10.1021/acs.jpclett.8b02980} {\bibfield
  {journal} {\bibinfo  {journal} {J. Phys. Chem Lett.}\ }\textbf {\bibinfo
  {volume} {9}},\ \bibinfo {pages} {6511} (\bibinfo {year} {2018})}\BibitemShut
  {NoStop}%
\bibitem [{\citenamefont {Kowalewski}\ \emph
  {et~al.}(2016{\natexlab{a}})\citenamefont {Kowalewski}, \citenamefont
  {Bennett},\ and\ \citenamefont {Mukamel}}]{Markus1}%
  \BibitemOpen
  \bibfield  {author} {\bibinfo {author} {\bibfnamefont {M.}~\bibnamefont
  {Kowalewski}}, \bibinfo {author} {\bibfnamefont {K.}~\bibnamefont {Bennett}},
  \ and\ \bibinfo {author} {\bibfnamefont {S.}~\bibnamefont {Mukamel}},\ }\href
  {\doibase 10.1021/acs.jpclett.6b00864} {\bibfield  {journal} {\bibinfo
  {journal} {J. Phys. Chem. Lett.}\ }\textbf {\bibinfo {volume} {7}},\ \bibinfo
  {pages} {2050} (\bibinfo {year} {2016}{\natexlab{a}})}\BibitemShut {NoStop}%
\bibitem [{\citenamefont {Kowalewski}\ \emph
  {et~al.}(2016{\natexlab{b}})\citenamefont {Kowalewski}, \citenamefont
  {Bennett},\ and\ \citenamefont {Mukamel}}]{Kowalewski16jcp}%
  \BibitemOpen
  \bibfield  {author} {\bibinfo {author} {\bibfnamefont {M.}~\bibnamefont
  {Kowalewski}}, \bibinfo {author} {\bibfnamefont {K.}~\bibnamefont {Bennett}},
  \ and\ \bibinfo {author} {\bibfnamefont {S.}~\bibnamefont {Mukamel}},\ }\href
  {\doibase 10.1063/1.4941053} {\bibfield  {journal} {\bibinfo  {journal} {J.
  Chem. Phys.}\ }\textbf {\bibinfo {volume} {144}},\ \bibinfo {pages} {054309+}
  (\bibinfo {year} {2016}{\natexlab{b}})},\ \Eprint
  {http://arxiv.org/abs/1601.03694} {arXiv:1601.03694} \BibitemShut {NoStop}%
\bibitem [{\citenamefont {Bennett}\ \emph {et~al.}(2016)\citenamefont
  {Bennett}, \citenamefont {Kowalewski},\ and\ \citenamefont
  {Mukamel}}]{Markus3}%
  \BibitemOpen
  \bibfield  {author} {\bibinfo {author} {\bibfnamefont {K.}~\bibnamefont
  {Bennett}}, \bibinfo {author} {\bibfnamefont {M.}~\bibnamefont {Kowalewski}},
  \ and\ \bibinfo {author} {\bibfnamefont {S.}~\bibnamefont {Mukamel}},\ }\href
  {\doibase 10.1039/C6FD00095A} {\bibfield  {journal} {\bibinfo  {journal}
  {Faraday Discuss.}\ }\textbf {\bibinfo {volume} {194}},\ \bibinfo {pages}
  {259} (\bibinfo {year} {2016})}\BibitemShut {NoStop}%
\bibitem [{\citenamefont {Csehi}\ \emph {et~al.}(2017)\citenamefont {Csehi},
  \citenamefont {Hal\'asz}, \citenamefont {Cederbaum},\ and\ \citenamefont
  {Vib\'ok}}]{Vibok1}%
  \BibitemOpen
  \bibfield  {author} {\bibinfo {author} {\bibfnamefont {A.}~\bibnamefont
  {Csehi}}, \bibinfo {author} {\bibfnamefont {G.~J.}\ \bibnamefont {Hal\'asz}},
  \bibinfo {author} {\bibfnamefont {L.~S.}\ \bibnamefont {Cederbaum}}, \ and\
  \bibinfo {author} {\bibfnamefont {A.}~\bibnamefont {Vib\'ok}},\ }\href
  {\doibase 10.1021/acs.jpclett.7b00413} {\bibfield  {journal} {\bibinfo
  {journal} {J. Phys. Chem. Lett.}\ }\textbf {\bibinfo {volume} {8}},\ \bibinfo
  {pages} {1624} (\bibinfo {year} {2017})}\BibitemShut {NoStop}%
\bibitem [{\citenamefont {Szidarovszky}\ \emph {et~al.}(2018)\citenamefont
  {Szidarovszky}, \citenamefont {Hal{\'{a}}sz}, \citenamefont
  {Cs{\'{a}}sz{\'{a}}r}, \citenamefont {Cederbaum},\ and\ \citenamefont
  {Vib{\'{o}}k}}]{Vibok2}%
  \BibitemOpen
  \bibfield  {author} {\bibinfo {author} {\bibfnamefont {T.}~\bibnamefont
  {Szidarovszky}}, \bibinfo {author} {\bibfnamefont {G.~J.}\ \bibnamefont
  {Hal{\'{a}}sz}}, \bibinfo {author} {\bibfnamefont {A.~G.}\ \bibnamefont
  {Cs{\'{a}}sz{\'{a}}r}}, \bibinfo {author} {\bibfnamefont {L.~S.}\
  \bibnamefont {Cederbaum}}, \ and\ \bibinfo {author} {\bibfnamefont
  {{\'{A}}.}~\bibnamefont {Vib{\'{o}}k}},\ }\href@noop {} {\bibfield  {journal}
  {\bibinfo  {journal} {J. Phys. Chem. Lett.}\ } (\bibinfo {year}
  {2018})}\BibitemShut {NoStop}%
\bibitem [{\citenamefont {Flick}\ \emph
  {et~al.}(2017{\natexlab{a}})\citenamefont {Flick}, \citenamefont
  {Ruggenthaler}, \citenamefont {Appel},\ and\ \citenamefont {Rubio}}]{Rubio1}%
  \BibitemOpen
  \bibfield  {author} {\bibinfo {author} {\bibfnamefont {J.}~\bibnamefont
  {Flick}}, \bibinfo {author} {\bibfnamefont {M.}~\bibnamefont {Ruggenthaler}},
  \bibinfo {author} {\bibfnamefont {H.}~\bibnamefont {Appel}}, \ and\ \bibinfo
  {author} {\bibfnamefont {A.}~\bibnamefont {Rubio}},\ }\href {\doibase
  10.1073/pnas.1615509114} {\bibfield  {journal} {\bibinfo  {journal} {Proc.
  Natl. Acad. Sci.}\ }\textbf {\bibinfo {volume} {114}},\ \bibinfo {pages}
  {3026} (\bibinfo {year} {2017}{\natexlab{a}})}\BibitemShut {NoStop}%
\bibitem [{\citenamefont {Flick}\ \emph
  {et~al.}(2017{\natexlab{b}})\citenamefont {Flick}, \citenamefont {Appel},
  \citenamefont {Ruggenthaler},\ and\ \citenamefont {Rubio}}]{Rubio2}%
  \BibitemOpen
  \bibfield  {author} {\bibinfo {author} {\bibfnamefont {J.}~\bibnamefont
  {Flick}}, \bibinfo {author} {\bibfnamefont {H.}~\bibnamefont {Appel}},
  \bibinfo {author} {\bibfnamefont {M.}~\bibnamefont {Ruggenthaler}}, \ and\
  \bibinfo {author} {\bibfnamefont {A.}~\bibnamefont {Rubio}},\ }\href
  {\doibase 10.1021/acs.jctc.6b01126} {\bibfield  {journal} {\bibinfo
  {journal} {J. Chem. Theory Comput.}\ }\textbf {\bibinfo {volume} {13}},\
  \bibinfo {pages} {1616} (\bibinfo {year} {2017}{\natexlab{b}})}\BibitemShut
  {NoStop}%
\bibitem [{\citenamefont {Ruggenthaler}\ \emph {et~al.}(2018)\citenamefont
  {Ruggenthaler}, \citenamefont {Tancogne-Dejean}, \citenamefont {Flick},
  \citenamefont {Appel},\ and\ \citenamefont {Rubio}}]{Rubio3}%
  \BibitemOpen
  \bibfield  {author} {\bibinfo {author} {\bibfnamefont {M.}~\bibnamefont
  {Ruggenthaler}}, \bibinfo {author} {\bibfnamefont {N.}~\bibnamefont
  {Tancogne-Dejean}}, \bibinfo {author} {\bibfnamefont {J.}~\bibnamefont
  {Flick}}, \bibinfo {author} {\bibfnamefont {H.}~\bibnamefont {Appel}}, \ and\
  \bibinfo {author} {\bibfnamefont {A.}~\bibnamefont {Rubio}},\ }\href
  {http://dx.doi.org/10.1038/s41570-018-0118} {\bibfield  {journal} {\bibinfo
  {journal} {Nat. Rev. Chem.}\ }\textbf {\bibinfo {volume} {2}},\ \bibinfo
  {pages} {0118} (\bibinfo {year} {2018})}\BibitemShut {NoStop}%
\bibitem [{\citenamefont {Flick}\ \emph {et~al.}(2018)\citenamefont {Flick},
  \citenamefont {Rivera},\ and\ \citenamefont {Narang}}]{Flick1}%
  \BibitemOpen
  \bibfield  {author} {\bibinfo {author} {\bibfnamefont {J.}~\bibnamefont
  {Flick}}, \bibinfo {author} {\bibfnamefont {N.}~\bibnamefont {Rivera}}, \
  and\ \bibinfo {author} {\bibfnamefont {P.}~\bibnamefont {Narang}},\ }\href
  {\doibase 10.1515/nanoph-2018-0067} {\bibfield  {journal} {\bibinfo
  {journal} {Nanophotonics}\ }\textbf {\bibinfo {volume} {7}},\ \bibinfo
  {pages} {1479} (\bibinfo {year} {2018})}\BibitemShut {NoStop}%
\bibitem [{\citenamefont {Triana}\ \emph {et~al.}(2018)\citenamefont {Triana},
  \citenamefont {Pel\'{a}ez},\ and\ \citenamefont
  {Sanz-Vicario}}]{Triana18jpca}%
  \BibitemOpen
  \bibfield  {author} {\bibinfo {author} {\bibfnamefont {J.~F.}\ \bibnamefont
  {Triana}}, \bibinfo {author} {\bibfnamefont {D.}~\bibnamefont {Pel\'{a}ez}},
  \ and\ \bibinfo {author} {\bibfnamefont {J.~L.}\ \bibnamefont
  {Sanz-Vicario}},\ }\href {\doibase 10.1021/acs.jpca.7b11833} {\bibfield
  {journal} {\bibinfo  {journal} {J. Phys. Chem. A}\ }\textbf {\bibinfo
  {volume} {122}},\ \bibinfo {pages} {2266} (\bibinfo {year}
  {2018})}\BibitemShut {NoStop}%
\bibitem [{\citenamefont {Vendrell}(2018{\natexlab{a}})}]{Oriol1}%
  \BibitemOpen
  \bibfield  {author} {\bibinfo {author} {\bibfnamefont {O.}~\bibnamefont
  {Vendrell}},\ }\href {\doibase
  https://doi.org/10.1016/j.chemphys.2018.02.008} {\bibfield  {journal}
  {\bibinfo  {journal} {Chem. Phys.}\ }\textbf {\bibinfo {volume} {509}},\
  \bibinfo {pages} {55 } (\bibinfo {year} {2018}{\natexlab{a}})}\BibitemShut
  {NoStop}%
\bibitem [{\citenamefont {Triana}\ and\ \citenamefont
  {Sanz-Vicario}(2019)}]{triana19prl}%
  \BibitemOpen
  \bibfield  {author} {\bibinfo {author} {\bibfnamefont {J.~F.}\ \bibnamefont
  {Triana}}\ and\ \bibinfo {author} {\bibfnamefont {J.~L.}\ \bibnamefont
  {Sanz-Vicario}},\ }\href@noop {} {\bibfield  {journal} {\bibinfo  {journal}
  {Phys. Rev. Lett.}\ }\textbf {\bibinfo {volume} {122}},\ \bibinfo {pages}
  {063603} (\bibinfo {year} {2019})}\BibitemShut {NoStop}%
\bibitem [{\citenamefont {Sun}\ \emph {et~al.}(2016)\citenamefont {Sun},
  \citenamefont {Wu}, \citenamefont {Ho},\ and\ \citenamefont
  {Rabitz}}]{Sun16arxiv}%
  \BibitemOpen
  \bibfield  {author} {\bibinfo {author} {\bibfnamefont {Q.}~\bibnamefont
  {Sun}}, \bibinfo {author} {\bibfnamefont {R.-B.}\ \bibnamefont {Wu}},
  \bibinfo {author} {\bibfnamefont {T.-S.}\ \bibnamefont {Ho}}, \ and\ \bibinfo
  {author} {\bibfnamefont {H.}~\bibnamefont {Rabitz}},\ }\href@noop {}
  {\enquote {\bibinfo {title} {Inherently trap-free convex landscapes for full
  quantum optimal control},}\ } (\bibinfo {year} {2016}),\ \Eprint
  {http://arxiv.org/abs/arXiv:1612.03988} {arXiv:1612.03988} \BibitemShut
  {NoStop}%
\bibitem [{\citenamefont {Gruebele}(2001)}]{Gruebel01cp}%
  \BibitemOpen
  \bibfield  {author} {\bibinfo {author} {\bibfnamefont {M.}~\bibnamefont
  {Gruebele}},\ }\href {\doibase https://doi.org/10.1016/S0301-0104(00)00404-3}
  {\bibfield  {journal} {\bibinfo  {journal} {Chem. Phys.}\ }\textbf {\bibinfo
  {volume} {267}},\ \bibinfo {pages} {33 } (\bibinfo {year}
  {2001})}\BibitemShut {NoStop}%
\bibitem [{\citenamefont {Shapiro}\ and\ \citenamefont
  {Brumer}(2011)}]{Shapiro11prl}%
  \BibitemOpen
  \bibfield  {author} {\bibinfo {author} {\bibfnamefont {M.}~\bibnamefont
  {Shapiro}}\ and\ \bibinfo {author} {\bibfnamefont {P.}~\bibnamefont
  {Brumer}},\ }\href {\doibase 10.1103/PhysRevLett.106.150501} {\bibfield
  {journal} {\bibinfo  {journal} {Phys. Rev. Lett.}\ }\textbf {\bibinfo
  {volume} {106}},\ \bibinfo {pages} {150501} (\bibinfo {year}
  {2011})}\BibitemShut {NoStop}%
\bibitem [{\citenamefont {Schlawin}\ and\ \citenamefont
  {Buchleitner}(2017)}]{Schlawin17njp}%
  \BibitemOpen
  \bibfield  {author} {\bibinfo {author} {\bibfnamefont {F.}~\bibnamefont
  {Schlawin}}\ and\ \bibinfo {author} {\bibfnamefont {A.}~\bibnamefont
  {Buchleitner}},\ }\href {\doibase 10.1088/1367-2630/aa55ec} {\bibfield
  {journal} {\bibinfo  {journal} {New J. Phys.}\ }\textbf {\bibinfo {volume}
  {19}},\ \bibinfo {pages} {013009} (\bibinfo {year} {2017})}\BibitemShut
  {NoStop}%
\bibitem [{\citenamefont {Rahav}\ and\ \citenamefont
  {Mukamel}(2010)}]{Rahav10pra}%
  \BibitemOpen
  \bibfield  {author} {\bibinfo {author} {\bibfnamefont {S.}~\bibnamefont
  {Rahav}}\ and\ \bibinfo {author} {\bibfnamefont {S.}~\bibnamefont
  {Mukamel}},\ }\href {\doibase 10.1103/PhysRevA.81.063810} {\bibfield
  {journal} {\bibinfo  {journal} {Phys. Rev. A}\ }\textbf {\bibinfo {volume}
  {81}},\ \bibinfo {pages} {063810} (\bibinfo {year} {2010})}\BibitemShut
  {NoStop}%
\bibitem [{\citenamefont {Dorfman}\ \emph {et~al.}(2016)\citenamefont
  {Dorfman}, \citenamefont {Schlawin},\ and\ \citenamefont
  {Mukamel}}]{Dorfman16rmp}%
  \BibitemOpen
  \bibfield  {author} {\bibinfo {author} {\bibfnamefont {K.~E.}\ \bibnamefont
  {Dorfman}}, \bibinfo {author} {\bibfnamefont {F.}~\bibnamefont {Schlawin}}, \
  and\ \bibinfo {author} {\bibfnamefont {S.}~\bibnamefont {Mukamel}},\ }\href
  {\doibase 10.1103/RevModPhys.88.045008} {\bibfield  {journal} {\bibinfo
  {journal} {Rev. Mod. Phys.}\ }\textbf {\bibinfo {volume} {88}},\ \bibinfo
  {pages} {045008} (\bibinfo {year} {2016})}\BibitemShut {NoStop}%
\bibitem [{\citenamefont {Kowalewski}\ \emph
  {et~al.}(2016{\natexlab{c}})\citenamefont {Kowalewski}, \citenamefont
  {Bennett},\ and\ \citenamefont {Mukamel}}]{Markus2}%
  \BibitemOpen
  \bibfield  {author} {\bibinfo {author} {\bibfnamefont {M.}~\bibnamefont
  {Kowalewski}}, \bibinfo {author} {\bibfnamefont {K.}~\bibnamefont {Bennett}},
  \ and\ \bibinfo {author} {\bibfnamefont {S.}~\bibnamefont {Mukamel}},\ }\href
  {\doibase 10.1063/1.4941053} {\bibfield  {journal} {\bibinfo  {journal} {J.
  Chem. Phys.}\ }\textbf {\bibinfo {volume} {144}},\ \bibinfo {pages} {054309+}
  (\bibinfo {year} {2016}{\natexlab{c}})}\BibitemShut {NoStop}%
\bibitem [{\citenamefont {Schleich}(2001)}]{Schleich}%
  \BibitemOpen
  \bibfield  {author} {\bibinfo {author} {\bibfnamefont {W.~P.}\ \bibnamefont
  {Schleich}},\ }\href
  {http://www.amazon.com/exec/obidos/redirect?tag=citeulike07-20\&path=ASIN/352729435X}
  {\emph {\bibinfo {title} {Quantum Optics in Phase Space}}},\ \bibinfo
  {edition} {1st}\ ed.\ (\bibinfo  {publisher} {Wiley-VCH},\ \bibinfo {year}
  {2001})\BibitemShut {NoStop}%
\bibitem [{\citenamefont {Meyer}\ \emph {et~al.}(1990)\citenamefont {Meyer},
  \citenamefont {Manthe},\ and\ \citenamefont {Cederbaum}}]{mctdh1}%
  \BibitemOpen
  \bibfield  {author} {\bibinfo {author} {\bibfnamefont {H.-D.}\ \bibnamefont
  {Meyer}}, \bibinfo {author} {\bibfnamefont {U.}~\bibnamefont {Manthe}}, \
  and\ \bibinfo {author} {\bibfnamefont {L.~S.}\ \bibnamefont {Cederbaum}},\
  }\href@noop {} {\bibfield  {journal} {\bibinfo  {journal} {Chem. Phys.
  Lett.}\ }\textbf {\bibinfo {volume} {165}},\ \bibinfo {pages} {73} (\bibinfo
  {year} {1990})}\BibitemShut {NoStop}%
\bibitem [{\citenamefont {Beck}\ \emph {et~al.}(2000)\citenamefont {Beck},
  \citenamefont {J$\ddot{a}$ckle}, \citenamefont {Worth},\ and\ \citenamefont
  {Meyer}}]{mctdh2}%
  \BibitemOpen
  \bibfield  {author} {\bibinfo {author} {\bibfnamefont {M.}~\bibnamefont
  {Beck}}, \bibinfo {author} {\bibfnamefont {A.}~\bibnamefont
  {J$\ddot{a}$ckle}}, \bibinfo {author} {\bibfnamefont {G.}~\bibnamefont
  {Worth}}, \ and\ \bibinfo {author} {\bibfnamefont {H.-D.}\ \bibnamefont
  {Meyer}},\ }\href@noop {} {\bibfield  {journal} {\bibinfo  {journal} {Phys.
  Rep.}\ }\textbf {\bibinfo {volume} {324}},\ \bibinfo {pages} {1} (\bibinfo
  {year} {2000})}\BibitemShut {NoStop}%
\bibitem [{\citenamefont {Werner}(2015)}]{molpro}%
  \BibitemOpen
  \bibfield  {author} {\bibinfo {author} {\bibfnamefont {H.-J.}\ \bibnamefont
  {Werner}},\ }\href {http://www.molpro.net} {\enquote {\bibinfo {title}
  {Molpro, version 2015.1, a package of ab initio programs.}}\ } (\bibinfo
  {year} {2015})\BibitemShut {NoStop}%
\bibitem [{\citenamefont {Gerry}\ and\ \citenamefont
  {Knight}(2005)}]{quantopt}%
  \BibitemOpen
  \bibfield  {author} {\bibinfo {author} {\bibfnamefont {C.}~\bibnamefont
  {Gerry}}\ and\ \bibinfo {author} {\bibfnamefont {P.}~\bibnamefont {Knight}},\
  }\href
  {https://www.cambridge.org/hu/academic/subjects/physics/optics-optoelectronics-and-photonics/introductory-quantum-optics?format=PB&isbn=9780521527354}
  {\emph {\bibinfo {title} {Introductory Quantum Optics}}},\ \bibinfo {edition}
  {1st}\ ed.\ (\bibinfo  {publisher} {Cambridge University Press},\ \bibinfo
  {year} {2005})\BibitemShut {NoStop}%
\bibitem [{\citenamefont {M$\o$ller}\ \emph {et~al.}(1996)\citenamefont
  {M$\o$ller}, \citenamefont {J$\o$rgensen},\ and\ \citenamefont
  {Dahl}}]{moller}%
  \BibitemOpen
  \bibfield  {author} {\bibinfo {author} {\bibfnamefont {K.~B.}\ \bibnamefont
  {M$\o$ller}}, \bibinfo {author} {\bibfnamefont {T.~G.}\ \bibnamefont
  {J$\o$rgensen}}, \ and\ \bibinfo {author} {\bibfnamefont {J.~P.}\
  \bibnamefont {Dahl}},\ }\href@noop {} {\bibfield  {journal} {\bibinfo
  {journal} {Phys. Rev. A}\ }\textbf {\bibinfo {volume} {54}},\ \bibinfo
  {pages} {5378} (\bibinfo {year} {1996})}\BibitemShut {NoStop}%
\bibitem [{\citenamefont {H{\'{e}}tet}\ \emph {et~al.}(2006)\citenamefont
  {H{\'{e}}tet}, \citenamefont {Gl\"ockl}, \citenamefont {Pilypas},
  \citenamefont {Harb}, \citenamefont {Buchler}, \citenamefont {Bachor},\ and\
  \citenamefont {Lam}}]{Hetet06jpb}%
  \BibitemOpen
  \bibfield  {author} {\bibinfo {author} {\bibfnamefont {G.}~\bibnamefont
  {H{\'{e}}tet}}, \bibinfo {author} {\bibfnamefont {O.}~\bibnamefont
  {Gl\"ockl}}, \bibinfo {author} {\bibfnamefont {K.~A.}\ \bibnamefont
  {Pilypas}}, \bibinfo {author} {\bibfnamefont {C.~C.}\ \bibnamefont {Harb}},
  \bibinfo {author} {\bibfnamefont {B.~C.}\ \bibnamefont {Buchler}}, \bibinfo
  {author} {\bibfnamefont {H.-A.}\ \bibnamefont {Bachor}}, \ and\ \bibinfo
  {author} {\bibfnamefont {P.~K.}\ \bibnamefont {Lam}},\ }\href {\doibase
  10.1088/0953-4075/40/1/020} {\bibfield  {journal} {\bibinfo  {journal} {J.
  Phys. B At. Mol. Opt.}\ }\textbf {\bibinfo {volume} {40}},\ \bibinfo {pages}
  {221} (\bibinfo {year} {2006})}\BibitemShut {NoStop}%
\bibitem [{\citenamefont {Ast}\ \emph {et~al.}(2013)\citenamefont {Ast},
  \citenamefont {Mehmet},\ and\ \citenamefont {Schnabel}}]{Ast13oe}%
  \BibitemOpen
  \bibfield  {author} {\bibinfo {author} {\bibfnamefont {S.}~\bibnamefont
  {Ast}}, \bibinfo {author} {\bibfnamefont {M.}~\bibnamefont {Mehmet}}, \ and\
  \bibinfo {author} {\bibfnamefont {R.}~\bibnamefont {Schnabel}},\ }\href
  {\doibase 10.1364/OE.21.013572} {\bibfield  {journal} {\bibinfo  {journal}
  {Opt. Express}\ }\textbf {\bibinfo {volume} {21}},\ \bibinfo {pages} {13572}
  (\bibinfo {year} {2013})}\BibitemShut {NoStop}%
\bibitem [{\citenamefont {Vendrell}(2018{\natexlab{b}})}]{Vendrell18prl}%
  \BibitemOpen
  \bibfield  {author} {\bibinfo {author} {\bibfnamefont {O.}~\bibnamefont
  {Vendrell}},\ }\href {\doibase 10.1103/physrevlett.121.253001} {\bibfield
  {journal} {\bibinfo  {journal} {Phys. Rev. Lett.}\ }\textbf {\bibinfo
  {volume} {121}},\ \bibinfo {pages} {253001+} (\bibinfo {year}
  {2018}{\natexlab{b}})}\BibitemShut {NoStop}%
\end{thebibliography}

\appendix
\section{Operators in Photon Displacement Coordinates}

The annihilation operator for a single mode is:
\begin{align}
a=\sqrt{\frac{\omega_c}{2\hbar}} \left( \hat{x} + \frac{i}{\omega_c} \hat{p} \right)
\end{align}
From that we can write the number operator in photon displacement coordinates:
\begin{align}
\hat n=a\da a = \frac{1}{2\hbar} \left( \omega_c \hat x^2  + \dfrac{\hat p^2}{\omega_c} -\dfrac{1}{2}\right)
\end{align}
which corresponds to $\hat H_c / \hbar \omega_c - 0.5$.
The expectation value of the photon number operator
is thus directly related to the energy expectation value of the mode:
\begin{align}
\ev{n} = \dfrac{\ev{H_c}}{\omega_c} - \dfrac{1}{2}
\end{align}

For the electric field we use the definition of the field operator:
\begin{align}
\hat E_c &= \dfrac{\varepsilon_c}{\sqrt{2}} \left( \hat a + \hat a\da \right)
\end{align}
which yields
\begin{align}\label{eq:Ecx}
\hat E_c = \sqrt{\dfrac{\omega_c}{\hbar}}\varepsilon_c \hat x
\end{align}

\end{document}